\documentclass{aastex631}

\usepackage{tabularx}
\usepackage{ulem}
\usepackage[thicklines]{cancel}
\usepackage{amsmath}

\begin{document}

\title{FAST Observations of the Microstructure in Interpulse Pulsars\footnote{Released on May, 2, 2025}}

\author[0009-0009-8247-3576]{Wei Li}
\affiliation{School of Physics and Electronic Science, Guizhou Normal University, Guiyang 550001, China \\}
\affiliation{Xinjiang Astronomical Observatory, Chinese Academy of Sciences, Urumqi, Xinjiang 830011, China\\}
\affiliation{University of Chinese Academy of Sciences, Beijing 100049, China\\}

\author[0000-0002-2060-5539]{Shijun Dang}
\email{dangsj@gznu.edu.cn (S.J. Dang)}
\affiliation{School of Physics and Electronic Science, Guizhou Normal University, Guiyang 550001, China \\}
\affiliation{Guizhou Provincial Key Laboratory of Radio Astronomy and Data Processing, Guizhou Normal University, Guiyang 550001, China\\}

\author[0000-0002-9786-8548]{Na Wang}
\email{na.wang@xao.ac.cn (N. Wang)}
\affiliation{Xinjiang Astronomical Observatory, Chinese Academy of Sciences, Urumqi, Xinjiang 830011, China\\}
\affiliation{University of Chinese Academy of Sciences, Beijing 100049, China\\}

\author{Chengmin Zhang}
\affiliation{National Astronomical Observatories, Chinese Academy of Sciences, Beijing 100101, China\\}

\author{Jingbo Wang}
\email{1983wangjingbo@163.com (J.B. Wang)}
\affiliation{Institute of Optoelectronic Technology, Lishui University, Lishui, Zhejiang 323000, China\\}

\author[0000-0002-5381-6498]{Jianping Yuan}
\affiliation{Xinjiang Astronomical Observatory, Chinese Academy of Sciences, Urumqi, Xinjiang 830011, China\\}

\author{Feifei Kou}
\affiliation{Xinjiang Astronomical Observatory, Chinese Academy of Sciences, Urumqi, Xinjiang 830011, China\\}

\author[0009-0009-9063-9118]{Yanqing Cai}
\affiliation{School of Physics and Electronic Science, Guizhou Normal University, Guiyang 550001, China \\}
\affiliation{University of Chinese Academy of Sciences, Beijing 100049, China\\}
\affiliation{National Astronomical Observatories, Chinese Academy of Sciences, Beijing 100101, China\\}

\author{Zurong Zhou}
\affiliation{National Time Service Center, Chinese Academy of Sciences, Xi’an 710600, China\\}
\affiliation{Key Laboratory of Time Reference and Applications, Chinese Academy of Sciences, Xi’an 710600, China\\}

\author{Shuangqiang Wang}
\affiliation{Xinjiang Astronomical Observatory, Chinese Academy of Sciences, Urumqi, Xinjiang 830011, China\\}

\author{Lunhua Shang}
\affiliation{School of Physics and Electronic Science, Guizhou Normal University, Guiyang 550001, China \\}
\affiliation{Guizhou Provincial Key Laboratory of Radio Astronomy and Data Processing, Guizhou Normal University, Guiyang 550001, China\\}

\author{Juntao Bai}
\affiliation{Xinjiang Astronomical Observatory, Chinese Academy of Sciences, Urumqi, Xinjiang 830011, China\\}

\author[0009-0006-4212-3801]{Yirong Wen}
\affiliation{Xinjiang Astronomical Observatory, Chinese Academy of Sciences, Urumqi, Xinjiang 830011, China\\}
\affiliation{University of Chinese Academy of Sciences, Beijing 100049, China\\}
\affiliation{Institute of Optoelectronic Technology, Lishui University, Lishui, Zhejiang 323000, China\\}

\author{Jing Zou}
\affiliation{Xinjiang Astronomical Observatory, Chinese Academy of Sciences, Urumqi, Xinjiang 830011, China\\}
\affiliation{University of Chinese Academy of Sciences, Beijing 100049, China\\}
\affiliation{Institute of Optoelectronic Technology, Lishui University, Lishui, Zhejiang 323000, China\\}

\author{Zhixiang Yu}
\affiliation{School of Physics and Electronic Science, Guizhou Normal University, Guiyang 550001, China \\}
\affiliation{Guizhou Provincial Key Laboratory of Radio Astronomy and Data Processing, Guizhou Normal University, Guiyang 550001, China\\}

\collaboration{20}{(AAS Journals Data Editors)}



\begin{abstract}
    In this study, we investigate the microstructure properties of four pulsars (PSRs J0953+0755 (B0950+08), J0627+0706, J0826+2637 (B0823+26) and J1946+1805 (B1944+17)) using the Five-hundred-meter Aperture Spherical radio Telescope (FAST), with particular emphasis on identifying microstructure within interpulse (IP). Through the application of autocorrelation function (ACF) analysis and fast Fourier transform (FFT) techniques, we have systematically examined the periodicity of microstructure in these pulsars. Our findings represent the first successful detection of microstructure within IP. Furthermore, we conducted a comprehensive statistical analysis comparing the characteristic timescales ($\tau_{\mu}$) and the characteristic periods $P_{\mu}$ of quasi-periodic microstructure between the main pulse (MP) and IP, and our results indicate that the $\tau_{\mu}$ and $P_{\mu}$ of microstructure across components appear consistent within measurement errors for PSR J0627+0706, but microstructure in IP are relatively smaller than those in MP for PSR J0953+0755. Furthermore, the relationship between $P_{\mu}$ of microstructure and the rotation period in neutron star populations was reconfirmed: $P_{\mu}(\text{ms})=(1.337\pm0.114)\times P(\text{s})^{(1.063\pm0.038)}$.
\end{abstract}



\section{Introduction} \label{sec:intro}
Pulsars are rapidly rotating neutron stars characterized by intense magnetic fields. Their pulse emissions exhibit significant variations from one period to another. Numerous intriguing single-pulse behaviors have been identified in pulsars, including mode changing, subpulse drifting, and nulling \citep{2007MNRAS.377.1383W}, among others. Investigating these single-pulse phenomena provides valuable insights into the magnetospheric structure and radiation mechanisms of pulsars.

Microstructure, which manifests as rapid intensity fluctuations within single-pulses on submillisecond timescales, represents another fascinating single-pulse phenomenon. Shortly after the initial discovery of pulsars, \cite{1968Natur.218.1122C} made the first observations of the microstructure in these celestial objects. Subsequent research has led to the detection of microstructure in an increasing number of pulsars, revealing their intriguing properties. In particular, microstructure has also been observed in the Crab pulsar \citep{2003Natur.422..141H, 2010A&A...524A..60J} and the Vela pulsar \citep{2001ApJ...549L.101J, 2002MNRAS.334..523K}, offering further opportunities to improve our understanding of these two prominent pulsars.

There is increasing evidence shows that certain microstructure exhibits quasiperiodic behavior \citep{1976ApJ...208L..43B, 1976Natur.260...25F, 1976ApJ...208..944C, 1977ApJ...218..484C, 1981IAUS...95..191B, 1981SvA....25..442S}. \cite{1971PhDT.........6H} first observed microstructure in PSR B0950+08 with $\tau_{\mu}$ shorter than $10\,\mu\text{s}$, and identified quasi-periodic microstructure with $P_{\mu}$ ranging from $300$ to $700\,\mu\text{s}$. Subsequently, \cite{1990AJ....100.1882C} detected quasi-periodic microstructure in several pulsars, including PSRs B0950+08, B1133+16, B0809+74, B1944+17 and B2016+28. Notably, quasi-periodic microstructure was also detected in a ultra-long-period pulsar J0901$-$4046 with a rotation period of approximately $76\,\text{s}$ \citep{2022NatAs...6..828C} and a 6.45-h-period coherent radio transient ASKAP J183950.5$-$075635.0 \citep{2025NatAs...9..393L}.

Interestingly, observational studies have revealed a potential connection between microstructure properties and the rotation period ($P$) of pulsars. \cite{1976ApJ...208..944C} established a relationship between the $\tau_{\mu}$ range of microstructure ($\bigtriangleup \tau_{\mu}$) and the pulsar's rotation period: $\bigtriangleup \tau_{\mu} \approx 10^{-3}P$. Furthermore, \cite{2002MNRAS.334..523K} derived a relationship between $\tau_{\mu}$ and $P$: $\tau _{\mu }(\mu\text{s})=(600\pm 100)P(\text{s})^{1.1\pm 0.2}$, emphasizing that this linear dependence is not an artifact of the time resolution of observational instruments. \cite{2015ApJ...806..236M} (hereafter M15) further demonstrated that $P_{\mu}$ tends to increase with the pulsar's rotation period, although no clear correlation was found between $P_{\mu}$ and other pulsar parameters, such as the period derivative, characteristic age, or surface magnetic field.


Efforts have also been made to detect microstructure in millisecond pulsars (MSPs), which require exceptionally high time resolution. If the empirical relationship between $\tau_{\mu}$ and $P$ for normal-period pulsars \citep{1976ApJ...208..944C} is extended to MSPs, the $\tau_{\mu}$ range of their microstructure would be less than $30\,\text{ns}$ for MSPs with rotation periods below $30\,\text{ms}$ (if we only divide the pulse into 1024 phase bins). This implies that observations must achieve a time resolution of at least $30\,\text{ns}$ to resolve the fine microstructure in MSPs. At a time resolution of $80\,\mu\text{s}$, \cite{1998ApJ...498..365J} found no evidence of microstructure in MSP J0437$-$4715. Similarly, \cite{2016MNRAS.463.3239L} reported no detection of microstructure in MSP 1713+0747 at a resolution of $140\,\text{ns}$. Finally, \cite{2016ApJ...833L..10D} made the first successful detection of microstructure in MSPs J0437$-$4715 and J2145$-$0750 at resolutions of $0.384\,\mu\text{s}$ and $0.96\,\mu\text{s}$, respectively. They found that the properties of microstructure in MSPs are analogous to those in normal-period pulsars, including quasi-periodic behavior. Furthermore, they established a relationship between $P_{\mu}$ and the rotation periods of MSPs: $P_{\mu }\left ( \mu\text{s} \right )=1.06\pm 0.63P\left ( \text{ms} \right )^{0.96\pm 0.09}$, which aligns with the relationship observed in normal-period pulsars. Subsequently, \cite{2022MNRAS.513.4037L} also identified microstructure in MSPs J1022+1001, J2145$-$0750, and J1744$-$1134, demonstrating that the relationship between $P_{\mu}$ and $P$ they derived is consistent with the earlier findings of \cite{2016ApJ...833L..10D}. Currently, studies on the microstructure of MSPs remain limited. Further observational research on microstructure in MSPs may enhance our understanding of the similarities and differences in the radiation mechanisms between MSPs and normal-period pulsars.

Here, we pose a thought-provoking question: is microstructure unique to radio pulsars, or does it also manifest in other radio sources? Evidence suggests that microstructure-like phenomena indeed exist in other radio sources. For instance, \cite{2022MNRAS.510.1996C} observed microstructure in the radio magnetar XTE J1810$-$197. Similarly, \cite{2021ApJ...919L...6M} identified a quasi-periodic structure with a period of approximately $2.3\,\mu$s in the Fast Radio Burst (FRB) 20200120E. Quasi-periodic structure was also detected in FRBs 20201020A \citep{2023AA...678A.149P}, 20201124A \citep{2022RAA....22l4004N}, 20191221A, 20210206A and 20210213A \citep{2022Natur.607..256C}.  Furthermore, quasi-periodic microstructure was also observed in Rotating Radio Transients (RRATs) using the FAST, including RRATs J1918$-$0449 \citep{2022ApJ...934...24C}, J0139+3336 \citep{2024MNRAS.528.1213D}, and J1913+1330 \citep{2024ApJ...972...59Z,2024MNRAS.527.4129Z,2025MNRAS.tmp..545T}. These observations intriguingly suggest a connection between pulsars, magnetars, RRATs, and FRBs through the common feature of quasi-periodic microstructure-like phenomena. Even more compelling, \cite{2024NatAs...8..230K} derived two key relationships among diverse radio sources: (1) the linear relationship between the rotation periods and the $\tau_{\mu}$ of quasi-periodic substructure: $P_{\mu}\,(\text{ms})=(0.59\pm 0.03)\times P\,(\text{s})^{(0.99\pm 0.02)}$, and (2) the linear relationship between the rotation periods and the characteristic periods:$\tau_{\mu}\,(\text{ms})=(0.94\pm 0.04)\times P\,(\text{s})^{(0.97\pm 0.05)}$. These relationships imply that the radiation mechanisms of these distinct radio sources may share underlying similarities. Consequently, in-depth studies of microstructure may be crucial for advancing our understanding of neutron star populations and even the origins and radiation mechanisms of FRBs.

Inter-pulse pulsars represent a unique subclass of pulsars characterized by emission profiles consisting of both a MP and an IP. The canonical lighthouse model of pulsars naturally predicts the occurrence of IP. According to this model, two emission beams are aligned along the open field lines of a dipolar magnetic field. Specifically, when the inclination angle $\alpha$ approaches 90 degrees (indicating an almost orthogonal rotation), observers can detect emission from both beams associated with two diametrically opposite magnetic poles. This configuration is known as the double-pole IP model. In this scenario, the two pulse components are separated by approximately 180 degrees in longitude, with no detectable low-level emission between the MP and IP. The duty cycles of each component are relatively short, typically accounting for only a few percent of the pulsar's period. An alternative explanation for IP generation is provided by the single-pole model \citep{1977MNRAS.181..761M}. This model assumes a small angle of inclination $\alpha$, which corresponds to a nearly aligned rotation. In this case, the pulse widths are significantly broader than those in the double-pole IP model, often spanning the entire or most of the pulsar period. Even when the two components are separated, a low-intensity emission bridge is typically observed between them \citep{2011MNRAS.414.1314M}.

In this study, we aim to detect and analyze the microstructure in both MP and IP of pulsars, with the goal of exploring the similarities and differences between microstructure in these two components. Such investigations may provide valuable insights into the radiation mechanisms of MP and IP. Using high-quality data from the FAST, we conduct a detailed analysis of the microstructure in four interpulse pulsars. The structure of this paper is organized as follows: Section \ref{sec:observation} describes the observational setup, data processing procedures. Section \ref{sec:method} provides the analysis methods. The results of our study are presented in Section \ref{sec:results}, while Sections \ref{sec:discussion} and \ref{sec:conclusion} offer a detailed discussion and a concise summary of our findings, respectively.

\section{Observation and Data Process} \label{sec:observation}
We acquired single-pulse data for interpulse pulsars from the publicly accessible FAST database. The observations were conducted using the L-band receiver of FAST, which covers a frequency range of $1050\,\text{MHz}$ to $1450\,\text{MHz}$. Detailed information regarding the pulsars in our sample and the observational parameters are summarized in Table \ref{Table:obs_table}.

\begin{table}[]
\caption{Observation Information of Four Interpulse Pulsars}
\label{Table:obs_table}
\centering
\begin{tabular}{ccccccc}
\hline
JName      & $P$   & $\tau_{c}$ & Obs Time     & $N$ & Time Resolution & $N_{bin}$\\
           & (ms) & (Myr)    & (yyyy/mm/dd) &     & ($\mu \text{s}$)        &          \\
\hline 
J0627+0706 & 0.475884937713(3)   & 0.253 & 2019/11/26 & 1257 & 49.152 & 4096\\
J0826+2637 & 0.53066051169(3)    & 4.92  & 2019/11/26 & 701  & 98.304 & 8192\\
J0953+0755 & 0.2530651649482(9)  & 17.5  & 2018/07/18 & 4678 & 49.152 & 4096\\
J1946+1805 & 0.4406184769108(8)  & 290   & 2019/12/04 & 1327 & 98.304 & 4096\\
\hline
\end{tabular}
\tablecomments{Columns 1$-$3 list the pulsar names, rotation periods, and characteristic ages, respectively, while columns 4$-$7 provide the observation time, the number of single pulses, the time resolution of observation, and the phase bins of the single pulses.}
\end{table}

The data processing involved the following steps. First, we applied the timing ephemeris from the ATNF pulsar catalog ({\sc PSRCAT-V2.1.1})\footnote{https://www.atnf.csiro.au/people/pulsar/psrcat/} \citep{2005AJ....129.1993M}. Subsequently, we folded the data using the {\sc DSPSR}\footnote{https://dspsr.sourceforge.net/} package \citep{2011PASA...28....1V} to produce single-pulse stacks with distinct phase bins per period for each pulsar. The phase bin details are provided in Table \ref{Table:obs_table}. Finally, we used the {\sc PAZ} and {\sc PAZI} plugins in the {\sc PSRCHIVE}\footnote{https://psrchive.sourceforge.net/} software to mitigate radio frequency interference (RFI) \citep{2004PASA...21..302H}. We first deleted data affected by narrow-band and impulsive RFI and 5\% of the band edges using {\sc PAZ} plugin. Then we used {\sc PAZI} plugin to check the pulse profiles and deleted frequency channels affected by RFI.

\begin{figure}[ht!]
\centering
\includegraphics[scale=0.8]{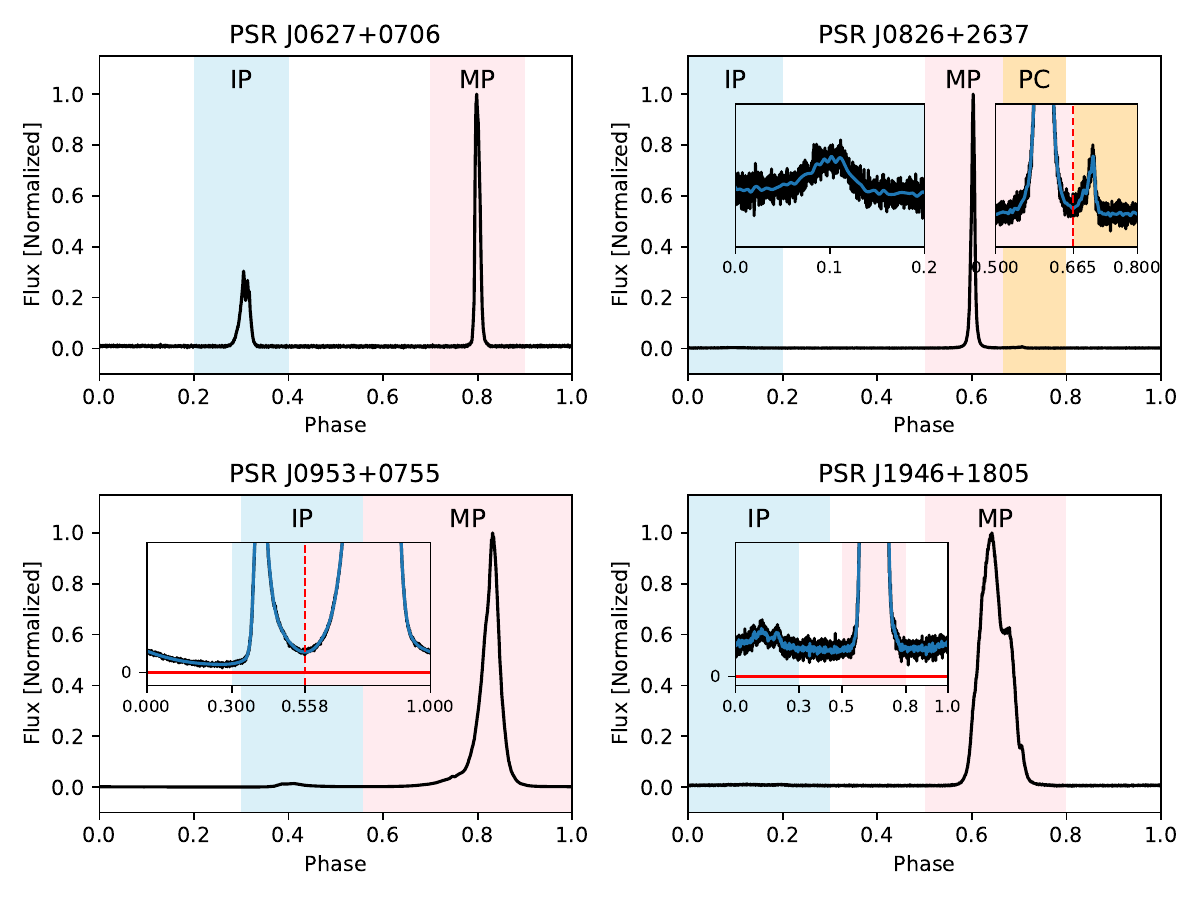}
\caption{The profiles of Pulsars. Each of the four panels illustrates the normalized integrated profile of a pulsar, including an inset that zooms in on the pulse at a specific phase range. The MP, IP, and PC components are annotated with pink, blue, and orange rectangles, respectively. Denoised profiles are represented by the blue lines. Vertical red lines in the insets of the top-right and bottom-left panels mark the divisions between adjacent components.
\label{fig:profile}}
\end{figure}

\begin{figure}[ht!]
\centering
\includegraphics[scale=0.8]{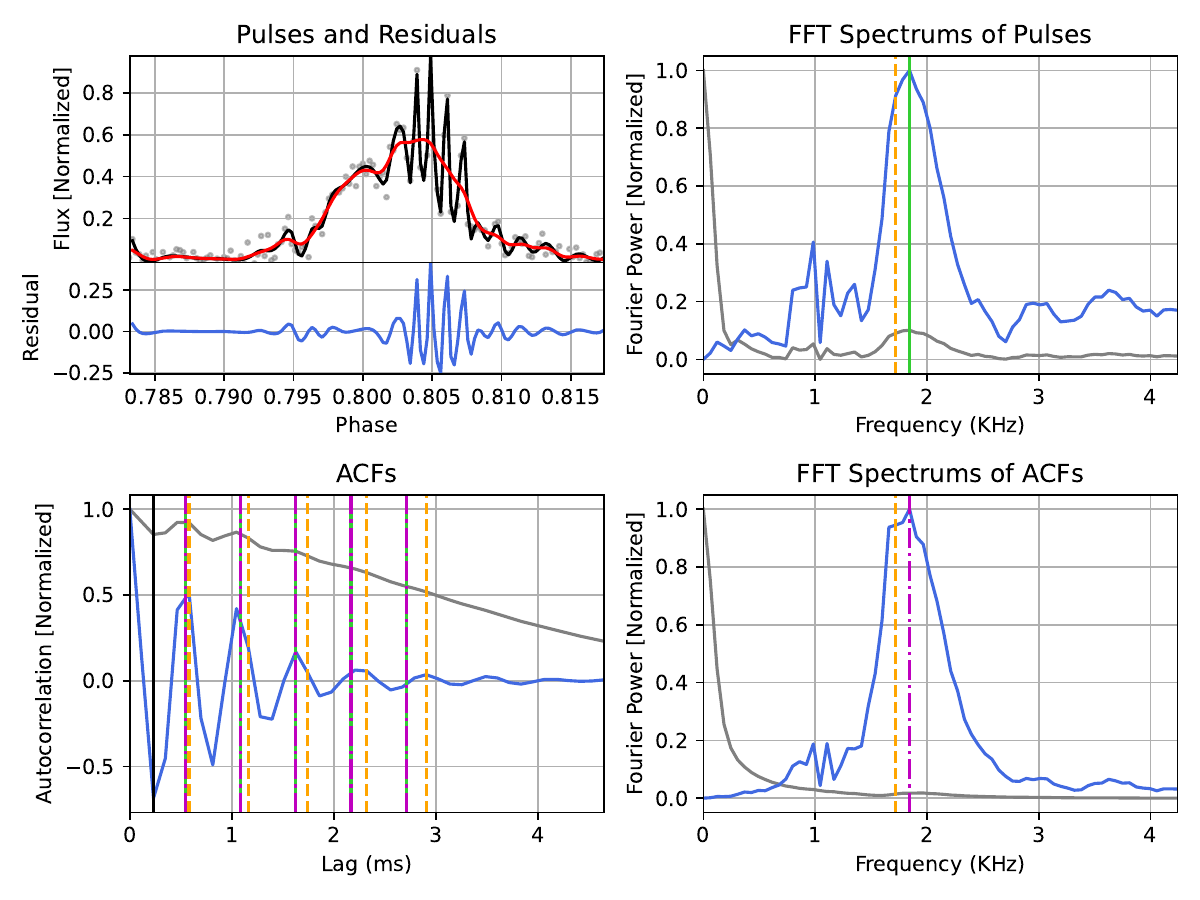}
\caption{Pulse 136: A representative example of quasi-periodic microstructure in MP of PSR J0627+0706. Top-left panel: The gray dots represent the original pulse profile, while the black line shows the denoised pulse. The red line corresponds to smoothed pulses obtained using smoothing bandwidths of $0.075\times N_{on}$. The residual of the smoothed pulse is plotted at the bottom. Lower-left panel: The blue line depict the normalized ACFs of the pulse residual. The solid black vertical line indicate the $\tau_{\mu}$ of the microstructure. The orange dashed vertical lines represent values of one to five times the $P_{\mu}$ of the quasi-periodic microstructure derived from the first method; the green solid vertical lines show values of one to five times $P_{\mu}$ obtained from the FFT of the pulse residual, and the magenta dashdot vertical lines correspond to values of one to five times $P_{\mu}$ calculated from the FFT of the ACF of the pulse residual.} Upper-right panel: The normalized FFT spectra of the denoised pulse and its residual. Lower-right panel: The normalized FFT spectra of the ACF of the denoised pulse and its residual.
\label{fig:0627+0706_mp_p136}
\end{figure}

\begin{figure}[ht!]
\centering
\includegraphics[scale=0.8]{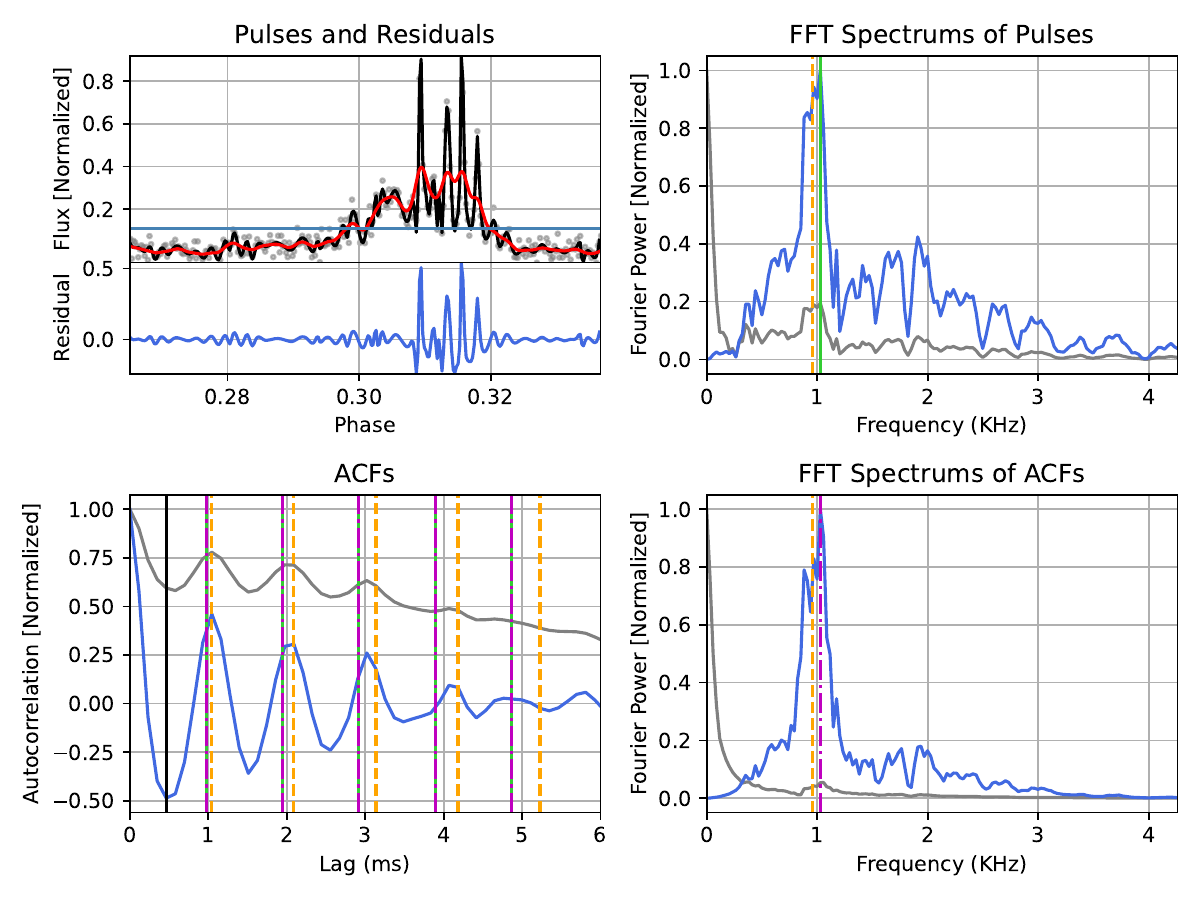}
\caption{Pulse 212: A representative example of quasi-periodic microstructure observed in the IP of PSR J0627+0706. The light blue horizontal solid line in the top-left panel represent three times the standard deviation of the off-pulse region. All other features and annotations follow the same conventions as described in Figure~\ref{fig:0627+0706_mp_p136}.
\label{fig:0627+0706_ip_p212}}
\end{figure}

\section{Methods} \label{sec:method}
After completing the data processing, we conducted an analysis of the microstructure in both the MP and IP. For determining the $\tau_{\mu}$ and $P_{\mu}$ of the microstructure, we primarily followed the method outlined in M15, albeit with some modifications. The key steps are as follows:

\begin{enumerate}
\item \textbf{Defining pulse components}
Prior to analyzing the microstructure within the profile components, we first established a reasonable delineation of these components, as illustrated in Figure \ref{fig:profile} and Table \ref{Table:component_table}. Since the profile components of PSRs J0627+0706 and J1946+1805 are distinctly separated, we were able to visually estimate the component boundarie. For PSR J0826+2637, the IP component is easily distinguishable, but the MP and post-cursor (PC) components are closely spaced. We reduced noise by applying a low-pass filter to the profile and identified the boundary between the two components as the phase of minimum intensity. This method was also used to separate the MP and IP components of PSR J0953+0755. While our delineation approach is not based on a rigorous definition, the subsequent findings confirm that it is entirely adequate for investigating the microstructure across different components.

\begin{table}[]
\caption{Delineation of Components in Pulsars}
\label{Table:component_table}
\centering
\begin{tabular}{cccc}
\hline
JName      & IP & MP & PC \\
           & & &  \\
\hline 
J0627+0706 & 0.2--0.4   & 0.7--0.9   & ---        \\
J0826+2637 & 0--0.2     & 0.5--0.665 & 0.665--0.8 \\
J0953+0755 & 0.3--0.558 & 0.558--1   & ---        \\
J1946+1805 & 0--0.3     & 0.5--0.8   & ---        \\
\hline
\end{tabular}
\tablecomments{Column 1 displays the pulsar names, and Columns 2 through 4 indicate the phase ranges of the IP, MP, and PC components, respectively.}
\end{table}

\item \textbf{Data screening:} To more precisely analyze the microstructure, we select pulses with a signal-to-noise ratio (SNR) greater than 15, where the SNR is defined as the ratio of the peak value within the on-pulse region to the root mean square (RMS) of the off-pulse region. Additionally, we require that at least five phase points within the on-pulse region exceed three times the standard deviation of the off-pulse region.

\item \textbf{Noise processing:} We employed smoothing spline regression, a non-parametric regression method, to separate potential signals from noise for each single pulse. This approach makes only assumptions about a true but unknown regression function without specifying its mathematical form in advance (see M15 for a detailed description of this method). Our results indicate that this method effectively removes noise while maintaining the intrinsic structure of the pulses, as illustrated in the top-left panels of Figures \ref{fig:0627+0706_mp_p136} and \ref{fig:0627+0706_ip_p212}. In these figures, the gray dots represent the original pulse profiles, while the black lines depict the denoised pulses.

\item \textbf{Removing low-frequency power and obtaining pulse residuals:} After completing the above steps, we initially attempted to calculate the ACF of the pulse directly to determine the $\tau_{\mu}$ of the microstructure and the $P_{\mu}$ of the quasi-periodic microstructure. However, this approach presents a challenge: the ACF of the pulse may be dominated by low-frequency power. To mitigate this issue and better reveal the periodicity of the microstructure, we aimed to remove the strong low-frequency components from the single pulse. Following the method in M15, we applied the Nadaraya-Watson kernel method to smooth the pulses (see M15 for details).

For the selection of the smoothing bandwidth $H$, which controls the degree of smoothness, we adopted a strategy similar to but not identical to that of M15. M15 noted that $H>0.1\times N_{on}$ resulted in overly smoothed pulses, while $H<0.05\times N_{on}$ caused the smoothed pulses to closely follow the denoised pulses. Due to the lack of a definitive criterion for selecting the optimal smoothing bandwidth, M15 used three distinct bandwidths $H$ (0.05$\times N_{on}$, 0.075$\times N_{on}$ and 0.1$\times N_{on}$) and analyzed the microstructure separately for each. Here, $N_{on}$ represents the number of phase bins of the on-pulse phase range of a single pulse.

In practice, we observed that the differences in results caused by varying smoothing bandwidths within the range of 0.05$\times N_{on}$ and 0.1$\times N_{on}$ are not significant, at least within the error margins displayed in M15. Therefore, we have directly chosen a smoothing bandwidth of 0.075$\times N_{on}$ for the smoothing process. After smoothing the denoised pulse, we derived the pulse residuals by subtracting the smoothed pulses from the denoised pulse (see the top-left panel in Figure \ref{fig:0627+0706_mp_p136}).

\item \textbf{Calculating ACFs:} We computed the ACFs of the pulse residuals. To facilitate a comparison between different ACFs, we also calculated the ACF of the denoised pulse and normalized all ACFs (see the lower-left panel in Figure \ref{fig:0627+0706_mp_p136}). 

\item \textbf{Determine $\tau_{\mu}$ of microstructure:} We selected the time delay corresponding to the first minimum of the ACF as the $\tau_{\mu}$ of the microstructure.

\item \textbf{Determine $P_{\mu}$ of microstructure:} Determining the $P_{\mu}$ of quasi-periodic microstructure is more challenging compared to determining the $\tau_{\mu}$. We employed three methods to determine $P_{\mu}$:

(1) We selected the time delay corresponding to the first minimum of the ACF as the $P_{\mu}$ of the microstructure.

(2) We performed a FFT on the pulse residual and used the reciprocal of the frequency corresponding to the peak in the FFT spectrum as a candidate value for $P_{\mu}$.

(3) We performed a FFT on the ACF of the pulse residual and used the reciprocal of the frequency corresponding to the peak in the FFT spectrum as $P_{\mu}$ of the pulse.

As illustrated in the lower-left panel of Figure \ref{fig:0627+0706_mp_p136}, the orange dashed vertical lines represent values of one to five times the $P_{\mu}$ of the quasi-periodic microstructure derived from the first method; the green solid vertical lines show values of one to five times $P_{\mu}$ obtained from the FFT of the pulse residual, and the magenta dashdot vertical lines correspond to values of one to five times $P_{\mu}$ calculated from the FFT of the ACF of the pulse residual. It should be noted that if the $P_{\mu}$ obtained by different methods are not significantly different, these lines will appear visually overlapping. In the upper-right panel of Figure \ref{fig:0627+0706_mp_p136}, we present the normalized FFT spectra of the denoised pulse and its residuals, while the lower-right panel displays the normalized FFT spectra of the ACFs of the denoised pulse and its residuals.

\item \textbf{Manual intervention}: During data analysis, it remains challenging for automated algorithms to reliably distinguish single pulses exhibiting obvious quasi-periodic characteristics or to determine the most appropriate pulse $P_{\mu}$ when selecting among three different analysis methods, often necessitating manual intervention for accurate judgment. Finally, we plotted the frequency-phase maps for all pulses in which quasi-periodic microstructure were detected, and further excluded the microstructure still affected by RFI through visual inspection.
\end{enumerate}

\section{Results} \label{sec:results}
\begin{figure}[ht!]
\centering
\includegraphics[scale=0.44]{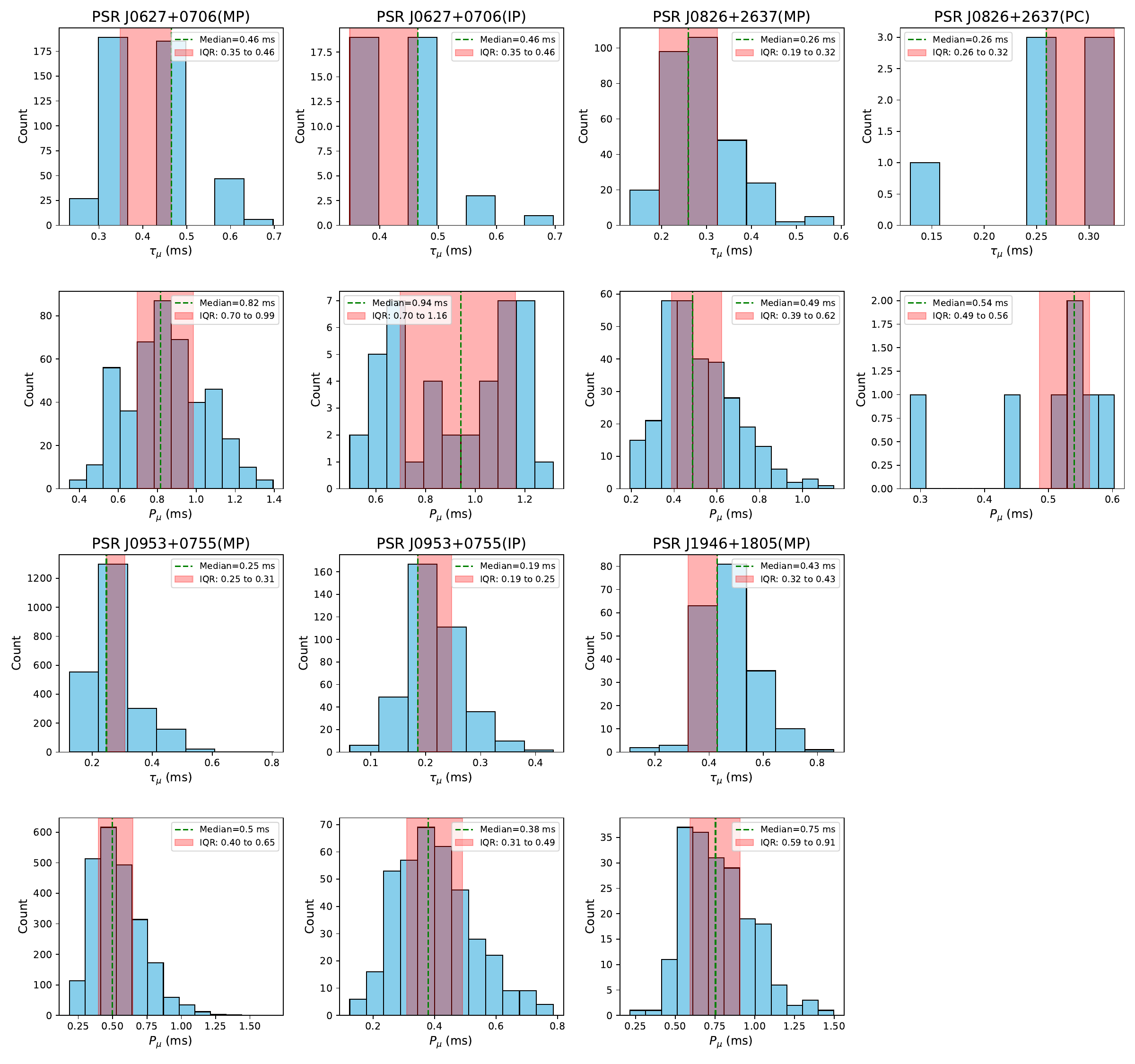}
\caption{Histograms of $\tau_{\mu}$ and $P_{\mu}$ for microstructure in four interpulse pulsars. For each pulsar component, the top panel displays the distribution of $\tau_{\mu}$ for quasi-periodic microstructure. And the bottom panel illustrates the distribution of $P_{\mu}$ for quasi-periodic microstructure. In each panel, the green dashed line indicates the median, while the pink shaded region represents the interquartile range (IQR).}
\label{fig:statistic}
\end{figure}

In this section, we present the results of our study on the microstructure in the four inter-pulse pulsars: PSRs J0627+0706, J0826+2637, J0953+0755, and J1946+1805. Using the methods outlined earlier, we determined the $\tau_{\mu}$ and $P_{\mu}$ of the microstructure in both MP and IP for each pulsar. The distributions of $\tau_{\mu}$ and $P_{\mu}$ are illustrated in Figure \ref{fig:statistic}. Detailed parameters of the microstructure in the MP and IP of these pulsars are provided in Table \ref{Table:micro_table}. Figure \ref{fig:comparison_mp_ip} shows the comparison of microstructure between MP and IP (or PC).

\begin{figure}[ht!]
\centering
\includegraphics[scale=0.8]{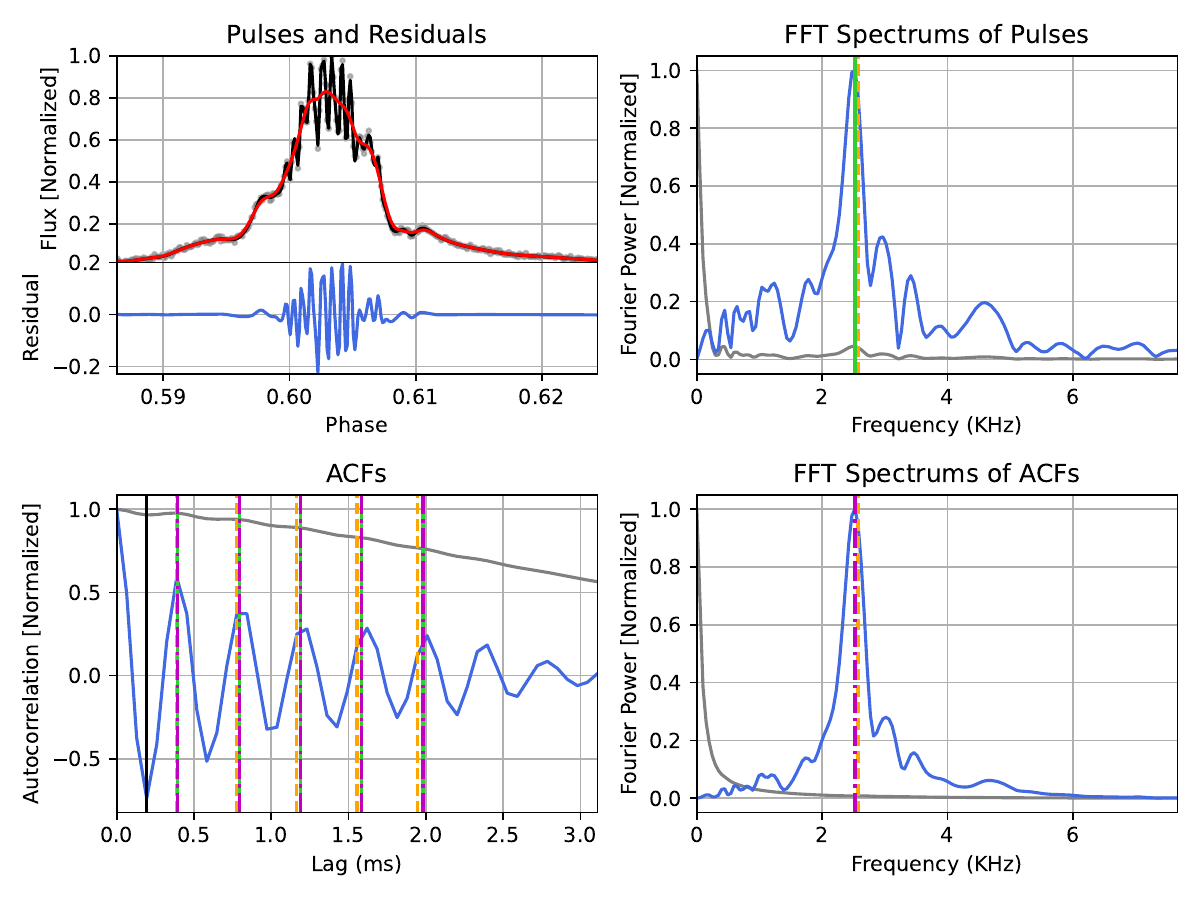}
\caption{Pulse 185: A representative example of quasi-periodic microstructure observed in the MP of PSR J0826+2637. All other features and annotations follow the same conventions as described in Figure~\ref{fig:0627+0706_mp_p136}.
\label{fig:0826+2637_mp_p185}}
\end{figure}
\begin{figure}[ht!]
\centering
\includegraphics[scale=0.8]{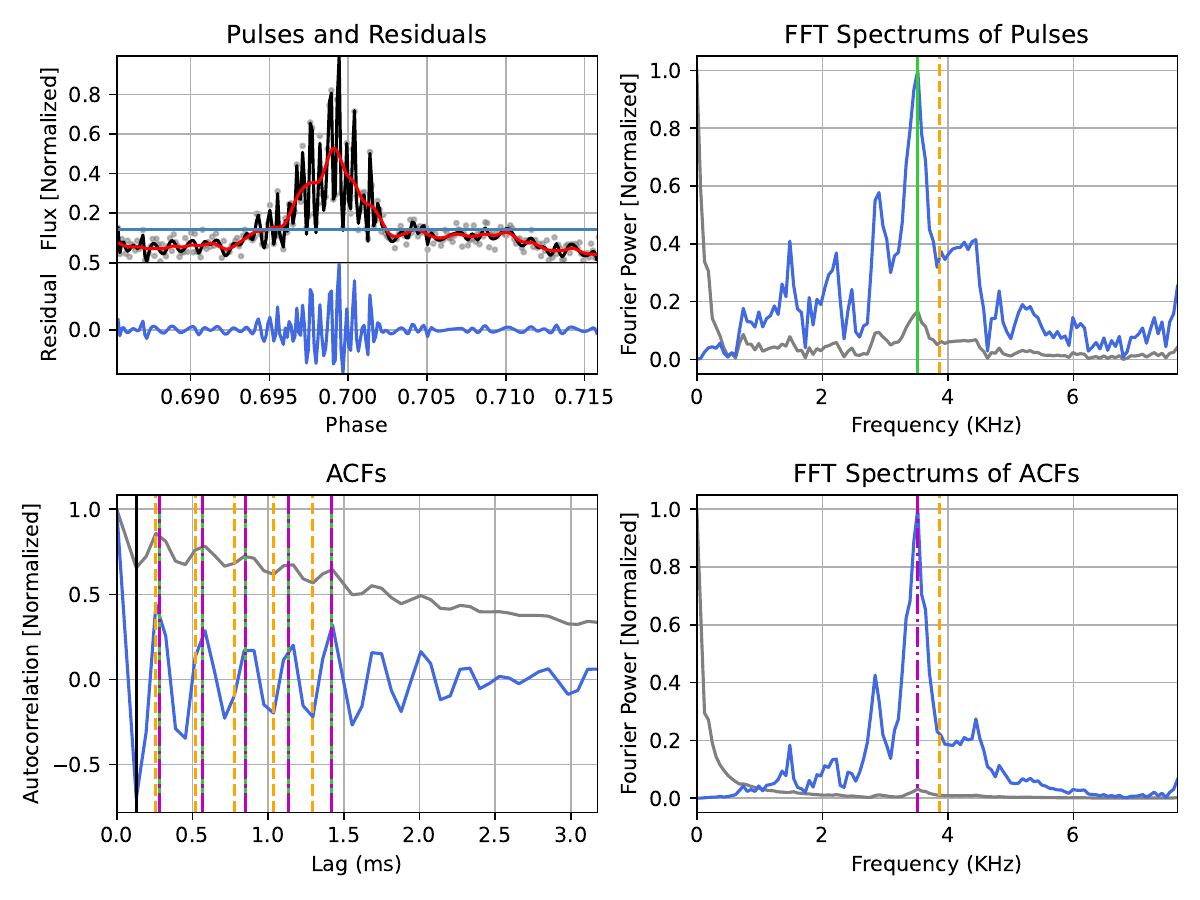}
\caption{Pulse 157: A representative example of quasi-periodic microstructure observed in the PC of PSR J0826+2637. The light blue horizontal solid line in the top-left panel represent three times the standard deviation of the off-pulse region. All other features and annotations follow the same conventions as described in Figure~\ref{fig:0627+0706_mp_p136}.
\label{fig:0826+2637_pc_p157}}
\end{figure}

\begin{table}[]
\caption{Microstructure Information of Four Interpulse Pulsars}
\label{Table:micro_table}
\centering
{
\begin{tabular}{ccccccc}
\hline
JName      & Component & $N$     & $N_{h}$ & $N_{m}$ & $\tau_{\mu}$            & $P_{\mu}$ \\
           &           &                          &         &         & (ms)                    & (ms)      \\
\hline 
J0627+0706 & MP        & 1257    & 1243    & 454     & $0.46^{+0.00}_{-0.23}$  & $0.82^{+0.17}_{-0.12}$  \\
J0627+0706 & IP        & 1257    & 792     & 42      & $0.46^{+0.00}_{-0.11}$  & $0.94^{+0.22}_{-0.24}$  \\
J0826+2637 & MP        & 701     & 648     & 303     & $0.26^{+0.06}_{-0.07}$  & $0.49^{+0.13}_{-0.10}$  \\
J0826+2637 & PC        & 701     & 9       & 7       & $0.26^{+0.06*}_{-0.00}$ & $0.54^{+0.02*}_{-0.05}$ \\
J0953+0755 & MP        & 4678    & 4651    & 2334    & $0.25^{+0.06}_{-0.00}$  & $0.50^{+0.15}_{-0.10}$  \\
J0953+0755 & IP        & 4678    & 1021    & 381     & $0.19^{+0.06}_{-0.00}$  & $0.38^{+0.11}_{-0.07}$  \\
J1946+1805 & MP        & 1327    & 454     & 95      & $0.43^{+0.00}_{-0.11}$  & $0.75^{+0.16}_{-0.16}$  \\
\hline
\end{tabular}
}
\tablecomments{Columns 1 and 2 list the pulsar names and the components of pulses, respectively. Columns 3, 4 and 5 provide the total number of pulses, the number of high signal-to-noise ratio pulses ($N_{h}$) and the number of pulses with quasi-periodic microstructure ($N_{m}$). Columns 6 and 7 present the $\tau_{\mu}$ and the $P_{\mu}$ of quasi-periodic microstructure. Additionally, data points marked with a superscript $*$ indicate results that lack statistical significance.}
\end{table}

\subsection{Timescale and Quasi-periodicity of Microstructure}\label{sec:scale_qp}
\subsubsection{PSR J0627+0706} \label{sec:0627+0706}
PSR J0627+0706 was initially identified in the Perseus Arm pulsar survey (currently unpublished), exhibiting a rotational period of approximately $0.476\,\text{s}$ \citep{2020MNRAS.494..228L} and a characteristic age $(\tau_{c}\approx P/(2 \dot P))$ of $0.253\,\text{Myr}$. According to \cite{2010MNRAS.402..745K}, this pulsar represents an orthogonal rotator, with its MP and IP separated by approximately $180^{\circ}$ in longitude, consistent with the upper-left panel in Figure \ref{fig:profile}. To date, no detailed investigation of the microstructure properties in this pulsar has been conducted.

In the MP component of this pulsar, our analysis revealed quasi-periodic microstructure in 454 pulses out of 1243 high signal-to-noise ratio pulses. Notably, the shortest observed $\tau_{\mu}$ in the MP was approximately $0.23\,\text{ms}$, which is comparable to twice the time resolution of our observations. This indicates that the current resolution may be insufficient to fully resolve the smallest microstructure in this pulsar. Furthermore, the minimum $P_{\mu}$ was measured to be $\sim 0.35\,\text{ms}$. A representative example of quasi-periodic microstructure (Pulse 136) is presented in Figure~\ref{fig:0627+0706_mp_p136}.

In the IP component of this pulsar, quasi-periodic microstructure was detected in 42 pulses out of 1243 high signal-to-noise ratio pulses. The shortest $\tau_{\mu}$ in the IP was measured to be approximately $0.35\,\text{ms}$, which is broader than the corresponding value in the MP ($0.23\,\text{ms}$). In addition, the minimum $P_{\mu}$ in the IP is about $0.49\,\text{ms}$, which also exceeds the smallest $P_{\mu}$ observed in the MP. A representative example of quasi-periodic microstructure in the IP (Pulse 212) is illustrated in Figure~\ref{fig:0627+0706_ip_p212}.

\subsubsection{PSR J0826+2637 (B0823+26)} \label{sec:826+2637}
PSR J0826+2637 (B0823+26) was first discovered by \cite{1968Natur.219.1237C}, exhibiting a rotation period of approximately $0.531\,\text{s}$ \citep{2004MNRAS.353.1311H} and a characteristic age of $\sim 4.92\,\text{Myr}$. The average pulse profile of this pulsar reveals a complex structure, including a MP, an IP, and a post-cursor component \citep{1973NPhS..243...77B}, which is in agreement with the lower-left panel in Figure \ref{fig:profile}. The IP and MP are separated by approximately $180^{\circ}$ in longitude, suggesting a nearly orthogonal emission geometry \citep{1986ApJ...304..256H}. This pulsar exhibits several notable emission properties, including mode-changing, nulling, and periodic fluctuation\citep{2015MNRAS.451.2493S,2018MNRAS.480.3655H, 2019MNRAS.487.4536B, 2023ApJ...946....2C}. The microstructure in this pulsar has been studied by \cite{1998A&A...332..111L} and M15, with \cite{1998A&A...332..111L} reporting that the $P_{\mu}$ of the microstructure range from $0.36\,\text{ms}$ to $0.66\,\text{ms}$.

In the MP component of this pulsar, quasi-periodic microstructure was detected in 303 pulses out of 648 high signal-to-noise ratio pulses. The shortest $\tau_{\mu}$ in the MP was approximately $0.13\,\text{ms}$, corresponding to about two times the time resolution of our observations. Additionally, the minimum $P_{\mu}$ was measured to be $0.19\,\text{ms}$. Our measured $P_{\mu} = 0.49^{+0.13}_{-0.10}\,\text{ms}$ for the MP component of this pulsar aligns with the previously reported range of $0.36\,\text{ms}$ to $0.66\,\text{ms}$ in \cite{1998A&A...332..111L}. A representative example of quasi-periodic microstructure (Pulse 185) is illustrated in Figure~\ref{fig:0826+2637_mp_p185}.

Notably, the signal from the IP component was too weak to resolve any microstructure. However, we detected quasi-periodic microstructure in the post-cursor (PC) component of this pulsar for the first time.

In the PC component, quasi-periodic microstructure was detected in only 7 pulses out of 9 high signal-to-noise ratio pulses. The shortest $\tau_{\mu}$ in the PC was found to be $0.13\,\text{ms}$, identical to the shortest $\tau_{\mu}$ in the MP. Moreover, the minimum $P_{\mu}$ in the PC was measured to be $0.28\,\text{ms}$, which is larger than that in the MP component. A representative example of quasi-periodic microstructure in the PC (Pulse 157) is shown in Figure~\ref{fig:0826+2637_pc_p157}.

Here, we emphasize that the statistical results for quasi-periodic microstructure in the PC component of this pulsar should be interpreted with caution due to the limited sample size (only 7 pulses), which may compromise their statistical significance. Additionally, the shortest observed $\tau_{\mu}$ in the MP and PC are only two times the time resolution of our observations, respectively, which suggests the potential existence of even finer microstructure in this pulsar that may not be resolvable with the current observational setup.

\begin{figure}[ht!]
\centering
\includegraphics[scale=0.8]{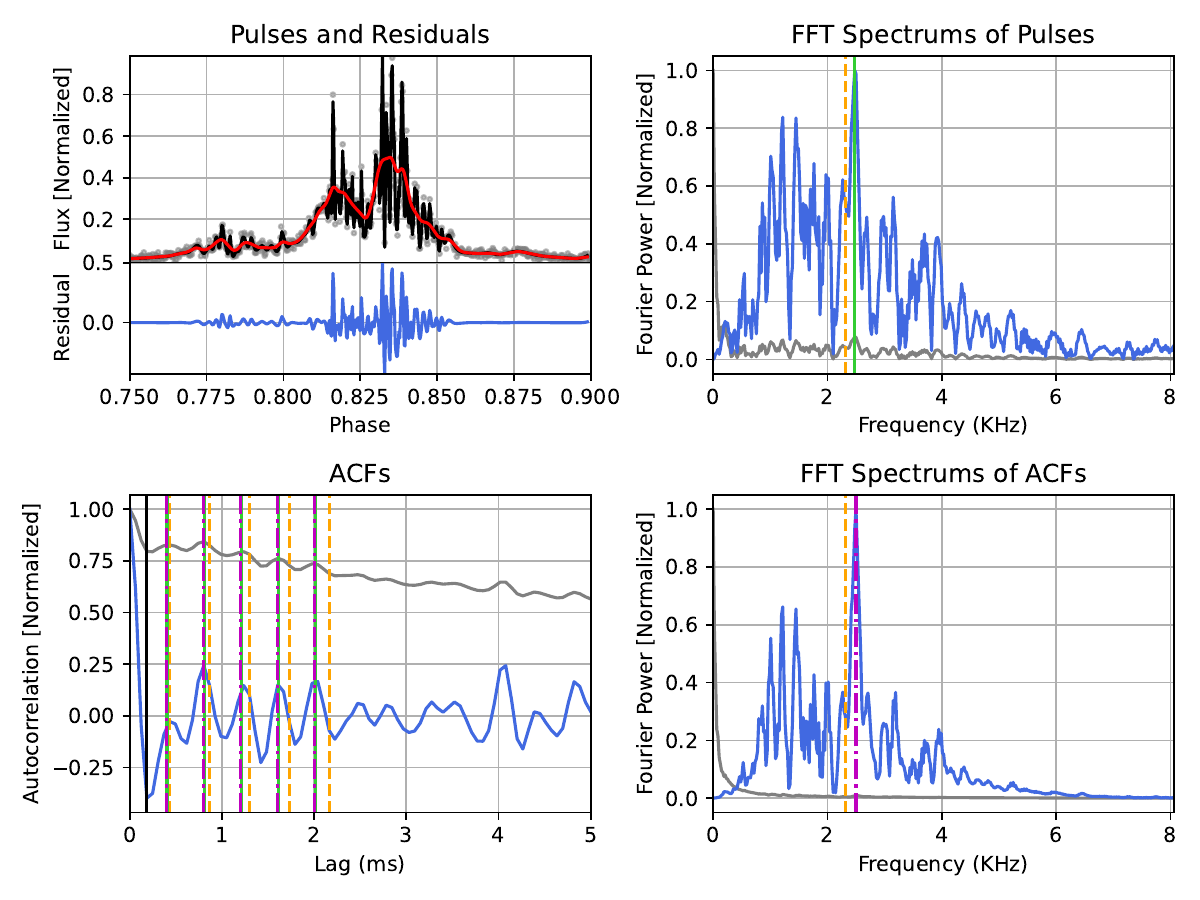}
\caption{Pulse 3357: A representative example of quasi-periodic microstructure observed in the MP of PSR J0953+0755. All other features and annotations follow the same conventions as described in Figure~\ref{fig:0627+0706_mp_p136}.
\label{fig:0953+0755_p3357}}
\end{figure}

\begin{figure}[ht!]
\centering
\includegraphics[scale=0.8]{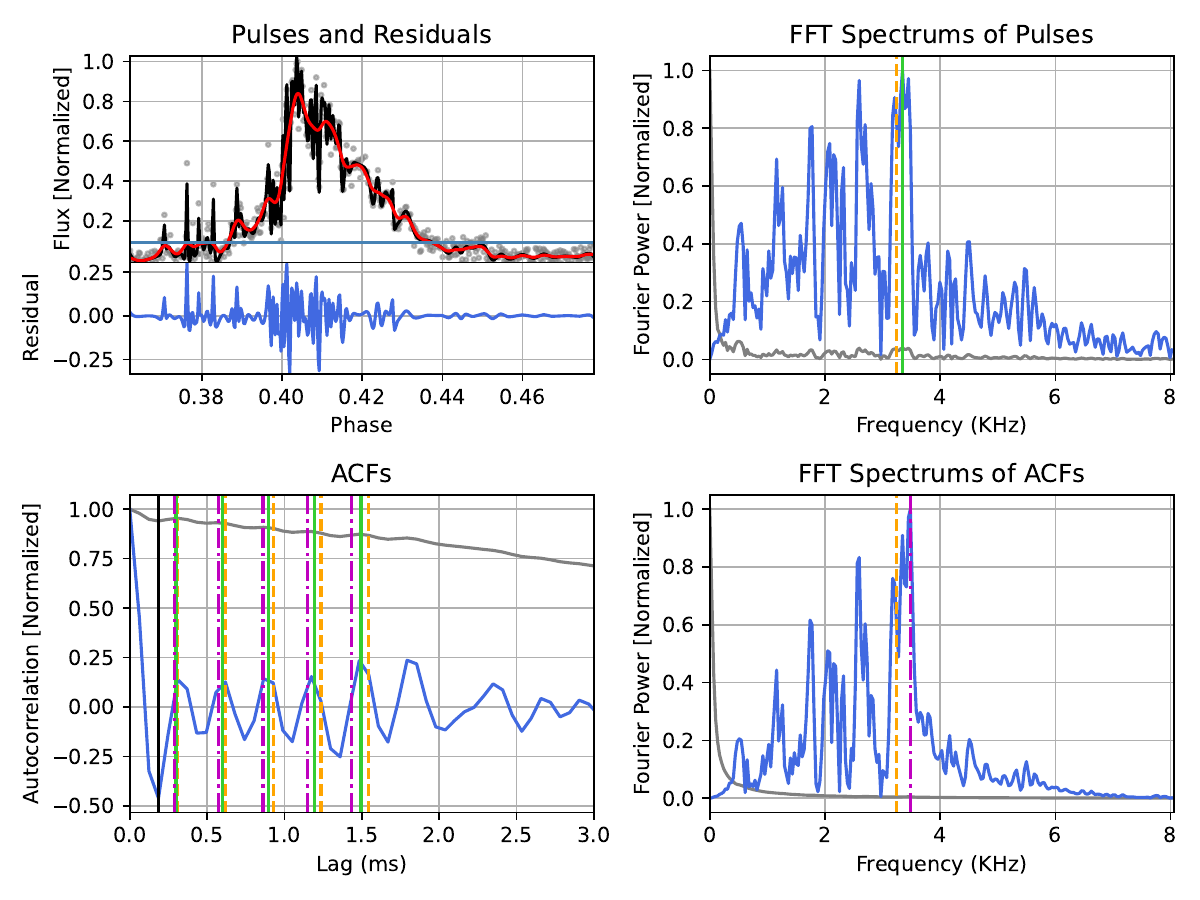}
\caption{Pulse 3070: A representative example showing quasi-periodic microstructure in IP of PSR J0953+0755. The light blue horizontal solid line in the top-left panel represent three times the standard deviation of the off-pulse region. All other features and annotations follow the same conventions as described in Figure~\ref{fig:0627+0706_mp_p136}.
\label{fig:0953+0755_ip_p3070}}
\end{figure}

\subsubsection{PSR J0953+0755 (B0950+08)} \label{sec:0953+0755}
PSR J0953+0755 (B0950+08) was first discovered by \cite{1968Natur.218..126P}, showing a rotation period of approximately $0.253\,\text{s}$ \citep{2004MNRAS.353.1311H} and a characteristic age of $\sim 17.5\,\text{Myr}$. The separation between its MP and IP remains frequency independent below $5\,\text{GHz}$ \citep{1981ApJ...249..241H}, and a bridge component connecting the IP and MP has been detected \citep{2023Univ....9...50Y}, which is in agreement with the upper-right panel in Figure \ref{fig:profile}. In particular, \cite{2022MNRAS.517.5560W} successfully detected signals from this pulsar over its entire rotation period using the FAST. The bridge emission and the emission over its entire rotation period indicate that this pulsar has a nearly-aligned emission geometry.

PSR B0950+08 is a well-studied pulsar known for its prominent microstructure features, as reported in numerous studies \citep{1968Natur.218.1122C,1971ApJ...169..487H,1978Natur.276...45H,1976ApJ...208..944C,1990AJ....100.1882C,1998A&A...332..111L,2002A&A...396..171P,2003MNRAS.344.1187K}. At $111.5\,\text{MHz}$, \cite{1971ApJ...169..487H} identified quasi-periodic microstructure with $P_{\mu}$ ranging from $0.3\,\text{ms}$ to $0.7\,\text{ms}$. Observations at $430\,\text{MHz}$ by \cite{1978Natur.276...45H} revealed that the $P_{\mu}$ of microstructure span $0.15\,\text{ms}$ to $1.1\,\text{ms}$, with the majority concentrated in the range of $0.4-0.6\,\text{ms}$. Furthermore, \cite{1998A&A...332..111L} measured the $\tau_{\mu}$ to be approximately $0.17\,\text{ms}$ at $4.85\,\text{GHz}$.

In the MP component of this pulsar, we detected 2334 pulses with quasi-periodic microstructure features out of 4651 high signal-to-noise ratio pulses. The shortest observed $\tau_{\mu}$ in the MP was approximately $0.06\,\text{ms}$, which is comparable to the time resolution of our observations. The minimum $P_{\mu}$ was measured to be $0.19\,\text{ms}$. Our measured $P_{\mu} = 0.50^{+0.15}_{-0.10}\,\text{ms}$ for the MP component of this pulsar aligns with the previously reported range of $0.3\,\text{ms}$ to $0.7\,\text{ms}$ in \cite{1971ApJ...169..487H}. A representative example of quasi-periodic microstructure (Pulse 3357) is illustrated in Figure~\ref{fig:0953+0755_p3357}.

In the IP component, quasi-periodic microstructure were detected in 381 pulses out of 1021 high signal-to-noise ratio pulses. The shortest $\tau_{\mu}$ in the IP was also found to be $0.06\,\text{ms}$, matching the shortest $\tau_{\mu}$ in the MP. The minimum $P_{\mu}$ in the IP was measured to be $0.12\,\text{ms}$, which is notably smaller than the corresponding value in the MP. A representative example of quasi-periodic microstructure in the IP (Pulse 3070) is shown in Figure~\ref{fig:0953+0755_ip_p3070}.

\begin{figure}[ht!]
\centering
\includegraphics[scale=0.8]{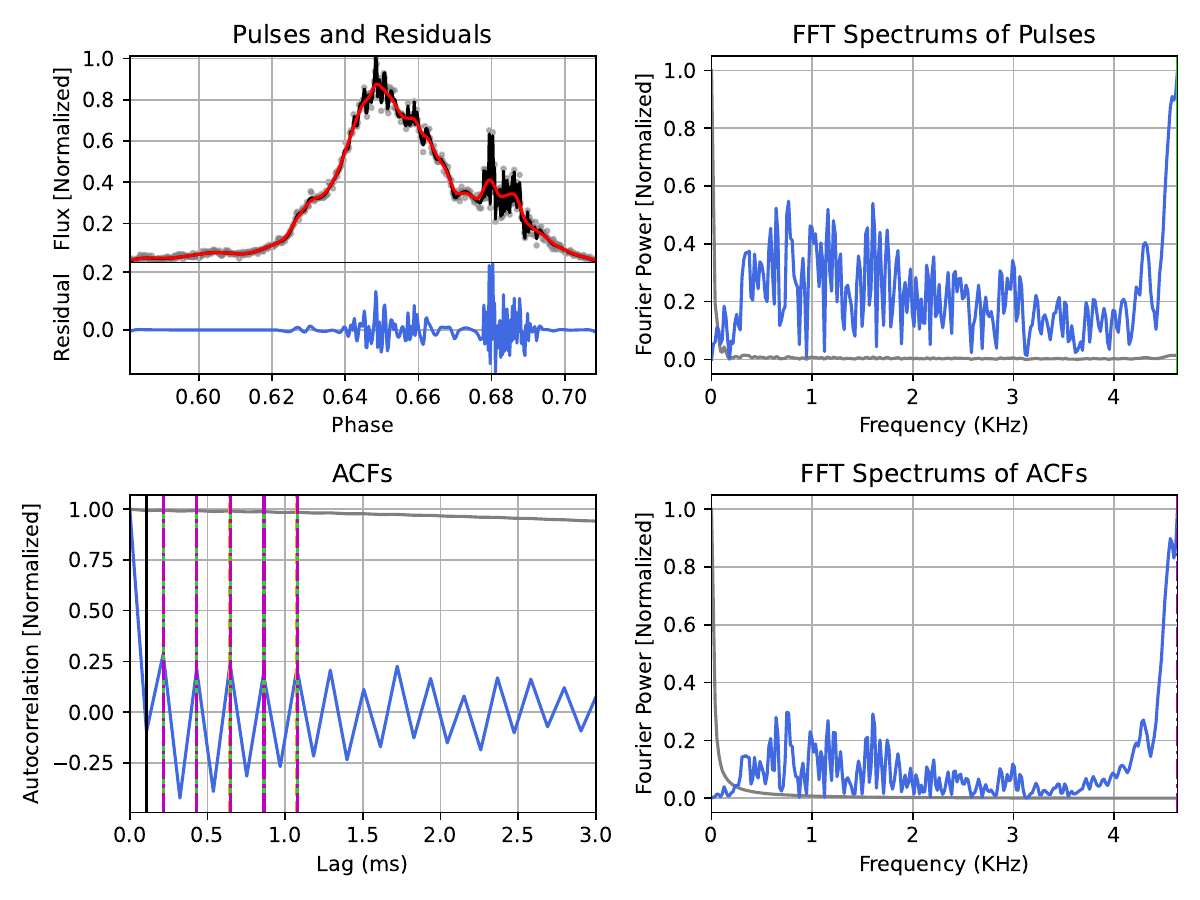}
\caption{Pulse 400: A representative example of quasi-periodic microstructure observed in the MP of PSR J1946+1805. All other features and annotations follow the same conventions as described in Figure~\ref{fig:0627+0706_mp_p136}.
\label{fig:1946+1805_mp_p400}}
\end{figure}

\subsubsection{PSR J1946+1805 (B1944+17)} \label{sec:1946+1805}
PSR J1946+1805 (B1944+17) was first discovered by \cite{1970Natur.225..167V}, exhibiting a rotation period of approximately $0.44\,\text{s}$ \citep{2004MNRAS.353.1311H} and a characteristic age of $\sim 290\,\text{Myr}$. The average pulse profile of this pulsar, as reported by \cite{1986ApJ...304..256H}, consists of a MP and an IP, with weak bridge emission detected between these two components, which is agreement with the lower-right panel in Figure \ref{fig:profile}. This emission pattern suggests an aligned rotator configuration, where the spin axis is nearly parallel to the magnetic axis, and both the MP and IP originate from the same magnetic pole. In addition to its pulse profile characteristics, this pulsar exhibits several notable emission properties, including nulling, subpulse drifting, and mode-changing \citep{1986ApJ...300..540D, 2010MNRAS.408...40K, 2024ApSS.369...21T}. Microstructure in this pulsar has been studied by \cite{1990AJ....100.1882C} and M15, with \cite{1990AJ....100.1882C} reporting the $P_{\mu}$ of quasi-periodic microstructure of approximately $0.8\,\text{ms}$.

In the MP component of this pulsar, quasi-periodic microstructure was detected in 95 pulses out of 454 high signal-to-noise ratio pulses. The shortest $\tau_{\mu}$ in the MP was approximately $0.11\,\text{ms}$, which matches the time resolution of our observations and suggests the potential existence of even finer microstructure that may not be resolvable with the current observational setup. The minimum $P_{\mu}$ was measured to be about $0.22\,\text{ms}$, corresponding to twice the time resolution of our observations. A representative example of quasi-periodic microstructure is illustrated in Figure~\ref{fig:1946+1805_mp_p400}.

Additionally, we attempted to investigate microstructure features in the IP component of this pulsar. However, as shown in Figure \ref{fig:profile}, due to the exceptionally weak intensity of the IP, no microstructure was detected.

\subsection{Comparison of Microstructure in MP and IP}
\begin{figure}[ht!]
\centering
\includegraphics[scale=0.8]{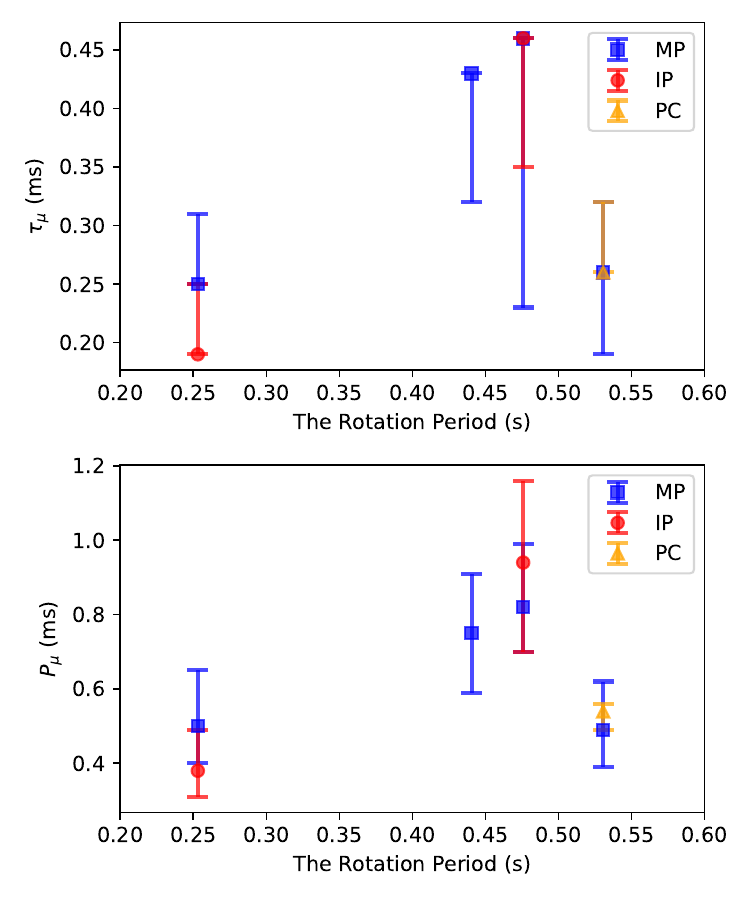}
\caption{The horizontal coordinate is the rotation period. The results of $\tau_{\mu}$ and $P_{\mu}$ of quasi-periodic microstructure are shown in the top panel and the bottom panel, respectively. The blue, the red and the orange dots represent the results of the MP, IP and PC components, respectively.
\label{fig:comparison_mp_ip}}
\end{figure}

In this subsection, we focus on comparing the similarities and differences between microstructure in MP and IP (PC).
In Figure \ref{fig:comparison_mp_ip}, the horizontal coordinate is the rotation period, and the results of $\tau_{\mu}$ and $P_{\mu}$ of quasi-periodic microstructure within are shown in the top panel and the bottom panel, respectively. The results demonstrate that the $\tau_{\mu}$ and $P_{\mu}$ of quasi-periodic microstructure across different pulse components show consistency within their respective error ranges. 

\begin{figure}[ht!]
\centering
\includegraphics[scale=0.8]{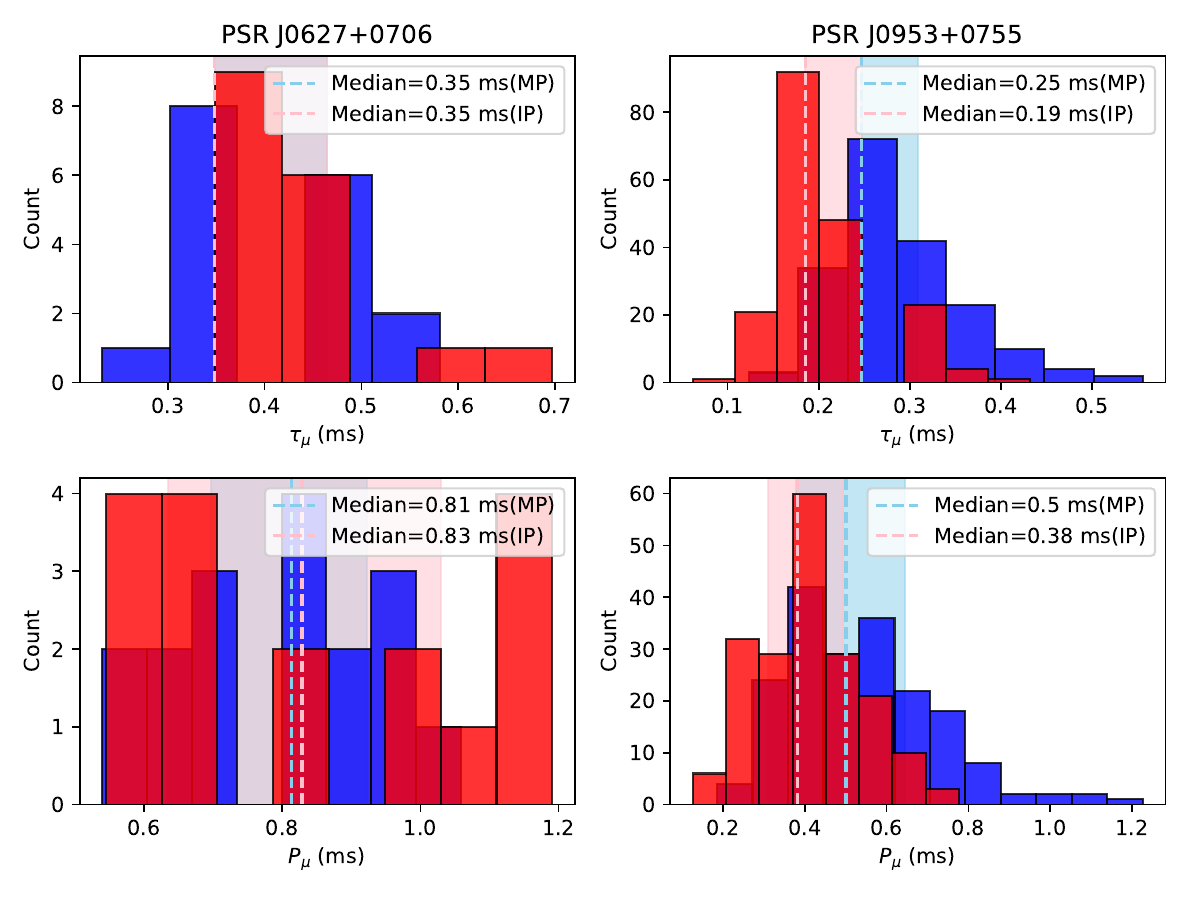}
\caption{Histograms of $\tau_{\mu}$ and $P_{\mu}$ for microstructure that occur simultaneously in single pulses in two interpulse pulsars. For each pulsar, the top panel displays the distribution of $\tau_{\mu}$ for quasi-periodic microstructure in MP (blue) and IP (or PC) (red), respectively. And the bottom panel illustrates the distribution of $P_{\mu}$ for these quasi-periodic microstructure. In each panel, the  dashed line indicates the median value, while the shaded region represents the interquartile range (IQR).}
\label{fig:comstats}
\end{figure}

To further confirm this result and explore the potential correlations between IP (PC) and different components of MP, we filtered out quasi-periodic microstructure that occur simultaneously in a single pulse. We detected the simultaneous occurrence of MP and IP (or PC) in 17, 3, and 190 single pulses for PSRs J0627+0706, J0826+2637 and J0953+0755, respectively. Additionally, we statistically analyzed the $\tau_{\mu}$ and $P_{\mu}$ distributions of these microstructure in PSRs J0627+0706 and J0953+0755. As shown in Figure \ref{fig:comstats}, the $\tau_{\mu}$ and $P_{\mu}$ of simultaneously occurring microstructure with different components in PSR J0627+0706 remain consistent within error margins. However, we noted that in PSR J0953+0755, both the $\tau_{\mu}$ and $P_{\mu}$ of microstructures in IP are relatively smaller than those in MP.

\begin{figure}[ht!]
\centering
\includegraphics[scale=0.8]{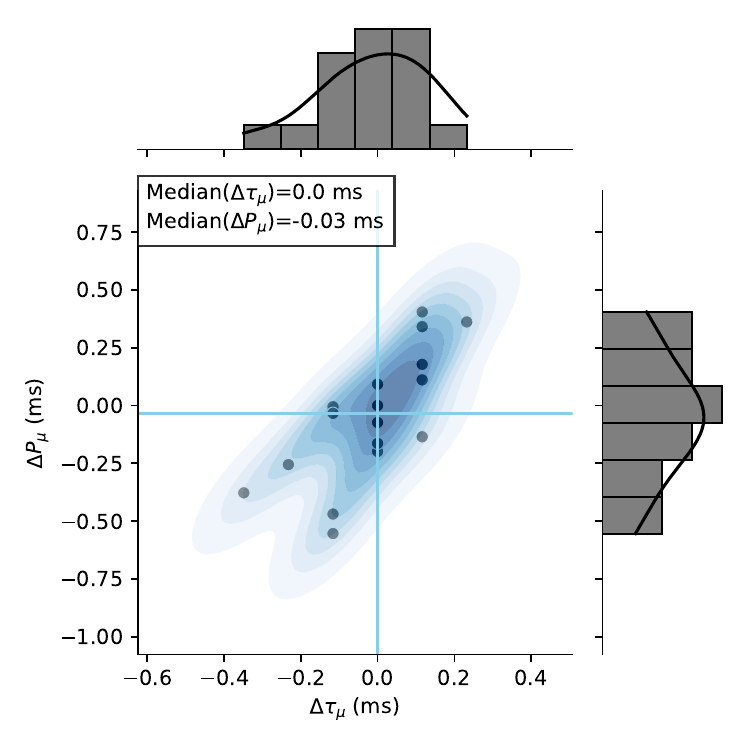}
\caption{The distribution of $\Delta\tau_{\mu}$ and $\Delta P_{\mu}$ in PSR J0627+0706. The black scatter points represent the distribution of $\Delta\tau_{\mu}$ and $\Delta P_{\mu}$, the blue shaded area is the two-dimensional kernel density estimation (KDE), and the sky-blue solid lines denote the medians of the of $\Delta\tau_{\mu}$ and $\Delta P_{\mu}$, respectively; the values in the text box at the upper left corner are the specific measured values of the corresponding medians (retained to two decimal places).
\label{fig:commicroJ0627}}
\end{figure}

\begin{figure}[ht!]
\centering
\includegraphics[scale=0.8]{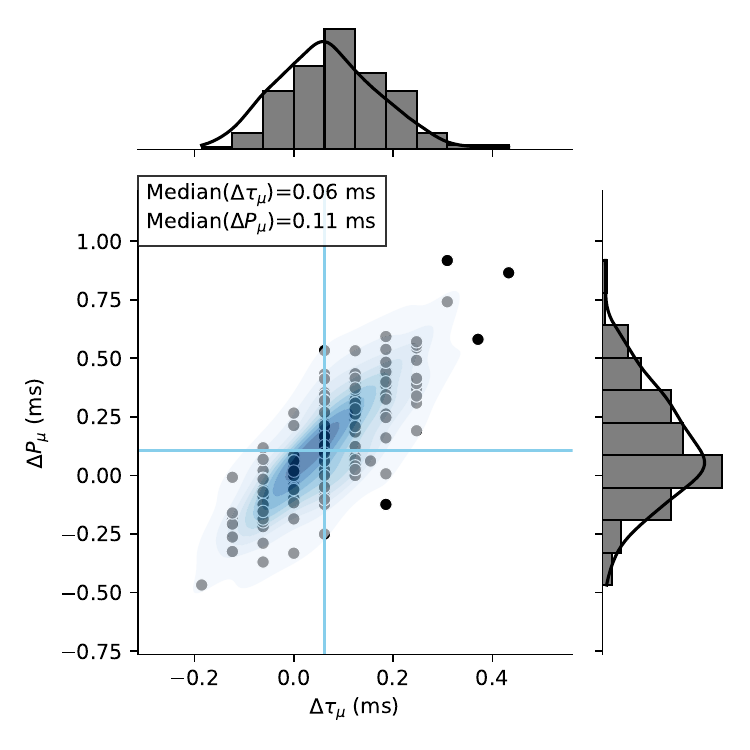}
\caption{The distribution of $\Delta\tau_{\mu}$ and $\Delta P_{\mu}$ in PSR J0953+0755. Others are the same with that in Figure \ref{fig:commicroJ0627}.
\label{fig:commicroJ0953}}
\end{figure}

Furthermore, to quantify the differences in $\tau_{\mu}$ and $P_{\mu}$ between simultaneously occurring microstructure of different components within the same sub-pulses, we defined $\Delta\tau_{\mu}=\tau_{\mu}^{MP}-\tau_{\mu}^{IP}$ and $\Delta P_{\mu}=P_{\mu}^{MP}-P_{\mu}^{IP}$. The two-dimensional distributions of parameters $\Delta\tau_{\mu}$ and $\Delta P_{\mu}$ for PSRs J0627+0706 and J0953+0755 are shown in Figures \ref{fig:commicroJ0627} and \ref{fig:commicroJ0953}, respectively. Kernel density estimates are overlaid on the scatter panels. Two blue solid lines (horizontal and vertical) represent the medians of $\Delta\tau_{\mu}$ and $\Delta P_{\mu}$, respectively. The results show that the differences in $\tau_{\mu}$ and $P_{\mu}$ between simultaneously occurring microstructure of MP and IP within the same sub-pulses of PSR J0627+0706 are not significant. 
However, these differences are relatively evident in PSR J0953+0755. It should be noted that the median of $\Delta\tau_{\mu}$ exactly corresponds to the time resolution of this pulsar, and the median of $\Delta P_{\mu}$ corresponds to twice the time resolution. Further investigation into this difference in the microstructure of this pulsar may require further improvement of the time resolution of observation.

Here, we have to point out that the small sample size and the limit of time resolution of observations may cause our results to be somewhat unrepresentative. Therefore, we look forward to more observations of the microstructure of interpulse pulsars, which will help us further understand the similarities and differences between microstructure in MP and IP (or PC).


\subsection{The Relationship between the Rotation Periods and Periods of Quasi-periodic Substructures}

\begin{figure}[ht!]
\centering
\includegraphics[scale=0.8]{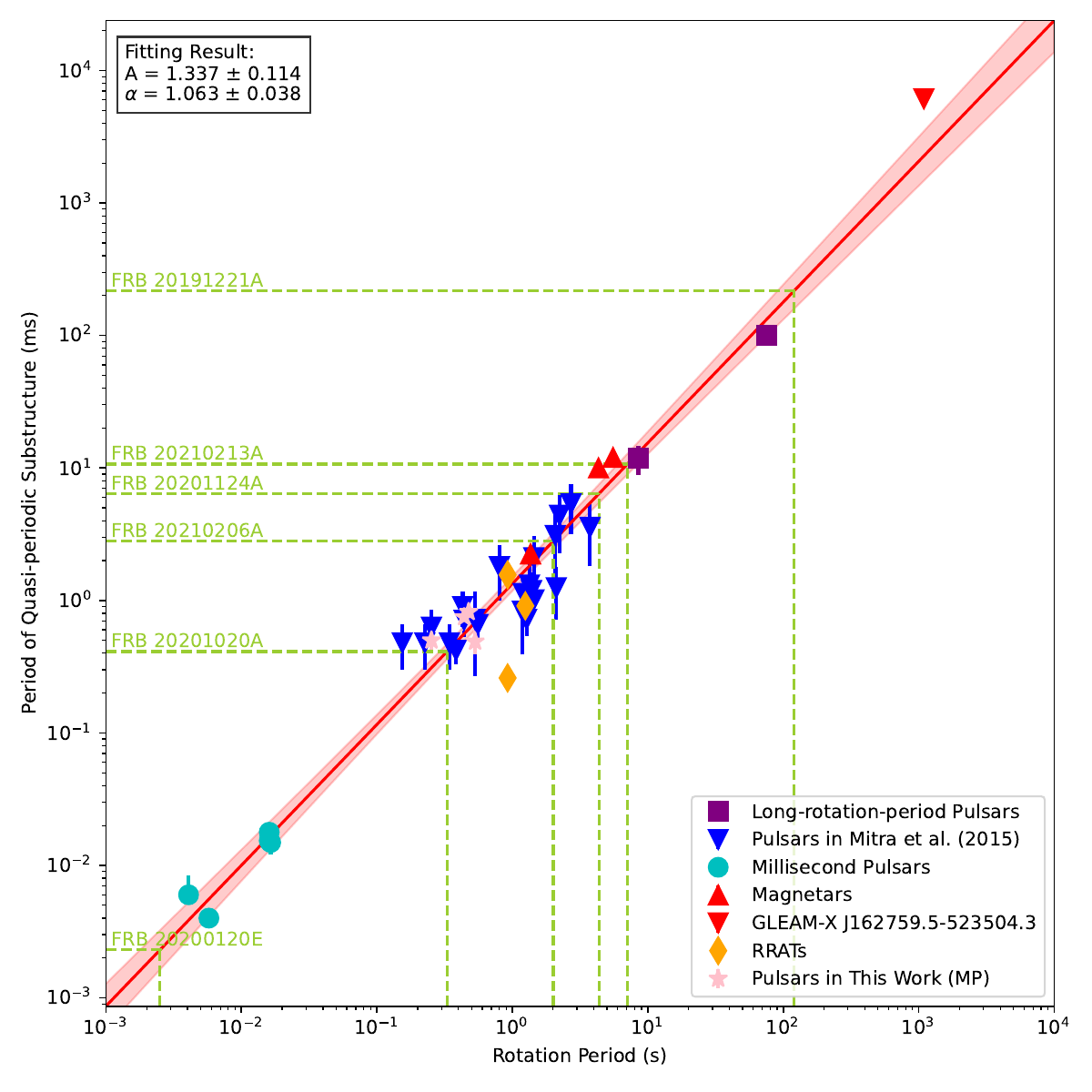}
\caption{The relationship between the rotation periods and periods of quasi-periodic substructures. A power function $P_{\mu}=A\times P^{\alpha}$ is used to fit these data points, where $P$ is in second. The fitting result is $P_{\mu}(\text{ms})=(1.337\pm0.114)\times P(\text{s})^{(1.063\pm0.038)}$. The error interval of $3\,\sigma$ is also shown in this figure. \label{fig:pp_relation}}
\end{figure}

The rotation period and the quasi-periodic substructure ($P-P_{\mu}$) relation of microstructures in neutron star populations(such as, normal pulsars, magnetars, long-rotation-period pulsars, MSPs and RRATs) has been widely studied\citep{2024NatAs...8..230K}. However, whether all pulsars follow this relation remains uncertain and requires investigation with a larger sample to be confirmed. By compiling previous data and incorporating new measurements from this work, we attempt to re-examine the $P-P_{\mu}$ relationship.

For magnetars, we cite the median of $P_{\mu}$ of magnetars in \citep{2024NatAs...8..230K}, including: PSR J1622$-$4950, XTE J1810$-$197, Swift J1818.0$-$1607 and GLEAM$-$X J162759.5$-$523504.3. For normal pulsars, data were mainly collected from M15 and this study. The data cited from M15 was obtained using smoothing bandwidths of $0.075\times N_{bin}$, and only that from the stronger integrated profile component was considered for the pulsars whose microstructure in different components were investigated (see M15 for details). Similarly, we only considered the microstructure data from the MP component in this work. For MSPs, we cite the results of all four MSPs in \citep{2016ApJ...833L..10D} and \citep{2022MNRAS.513.4037L}, including MSPs J0437$-$4715, J2145$-$0750, J1022+1001 and J1744$-$1134. For long-rotation-period pulsars, we cite the results of PSRs J1744$-$1134 \citep{2020MNRAS.492.2468M} and J0901-4046 \citep{2022NatAs...6..828C}. \cite{2022NatAs...6..828C} detected the quasi-periodic microstructure in PSR J0901$-$4046 with the rotation period of about $76\,\text{s}$, and reported the median $\tau_{\mu}=49\,\text{ms}$, the error of $\tau_{\mu}$ about $11\,\text{ms}$ and the relationship between $P_{\mu}$ and $\tau_{\mu}$: $(\tau_{\mu}=0.3782\times P_{\mu}+11.07$), but they did not provide the median of $P_{\mu}$. So we can only give a rough estimate of $P_{\mu}$ from the above relation and the median of $\tau_{\mu}$: $P_{\mu}=100\pm29\,\text{ms}$. Moreover, As far as we know, microstructure in only three RRATs have been studied, but we just use only three data from these RRATs for fitting \citep{2024MNRAS.528.1213D,2024MNRAS.527.4129Z,2025MNRAS.tmp..545T}, because \cite{2022ApJ...934...24C} only provided the $P_{\mu}$ ($\approx 2.31\pm0.25\,\text{ms}$) of an example of the quasi-periodic microstructure of RRAT J1918$–$0449, without statistical results of all quasi-periodic microstructure. \cite{2024MNRAS.528.1213D} gave the median $P_{\mu}$ ($\approx 0.91\,\text{ms}$) for the quasi-periodic microstructure in RRAT J0139+3336. \cite{2024MNRAS.527.4129Z} provided the mean value of $P_{\mu}\approx 1.56\,\text{ms}$ in RRAT J1913+1330. Based on the study by \cite{2025MNRAS.tmp..545T}, we determined a median $P_{\mu}$ of $0.26\,\text{ms}$ for the 12 quasi-periodic microstructure, with the interquartile range ($0.25-0.30\,\text{ms}$) representing the measurement uncertainty. Additionally, we noticed the quasi-periodic microstructure in RRAT J1913+1330 in \cite{2024ApJ...972...59Z}, although they did not conduct an in-depth analysis of the quasi-periodicity of microstructure. In addition, periods of these RRATs were cited from the RRATalog\footnote{https://rratalog.github.io/rratalog/}. In summary, our dataset for fitting includes magnetars, normal pulsars, long-rotation-period pulsars,MSPs and RRATs, as shown in Table \ref{Table:dataset_table}.

Due to the significant differences in magnitude spanning multiple orders of magnitude within the data, directly applying the least squares method to fit the power function model $P_{\mu} = A\times P^{\alpha}$ might lead to substantial errors. Therefore, we adopt a logarithmic transformation approach for processing: First, we perform a logarithmic conversion on the original data to obtain $\ln(P)$ and $\ln(P_{\mu})$. Then, based on the linearized model $\ln(P_{\mu}) = \ln(A) + \alpha\ln(P)$, we utilize the least squares method to conduct a linear fit on the transformed data. Finally, we convert the fitting results back to the original form of the power function. During this process, the errors of the model parameters are precisely calculated through the error propagation formula to ensure the accuracy and reliability of the results. 
As shown in Figure \ref{fig:pp_relation}, we confirm a linear dependency on rotational period: $P_{\mu}(\text{ms})=(1.337\pm0.114)\times P(\text{s})^{(1.063\pm0.038)}$.

In addition, We have noted some reports on quasi-periodic substructures in FRBs in recent years, including FRBs 20200120E \citep{2021ApJ...919L...6M}, 20191221A, 20210206A, 20210213A \citep{2022Natur.607..256C}, 20201020A \citep{2023AA...678A.149P} and 20201124A \citep{2022RAA....22l4004N} (see Table \ref{Table:frb_table} for details). In recent years, numerous theoretical models have been proposed to explain the radiation mechanisms of FRBs, while an increasing number of studies have focused on identifying their potential host stars \citep{2020Natur.587...59B, 2020Natur.587...54C}. Magnetars are leading candidate sources of FRBs \citep{2014MNRAS.442L...9L, 2017ApJ...843L..26B}. As shown in Figure \ref{fig:pp_relation}, the quasi-periodic substructure within magnetars exhibits an approximately the linear relationship between $P_{\mu}$ and $P$. If the radiation mechanism of the quasi-periodic substructures in FRBs is similar to that in other neutron stars, or if FRBs indeed originate from magnetars, then according to the periods of the quasi-periodic substructures in FRBs and our fitting results, we can roughly predict the spin periods of the FRBs' host stars ($P_{F}$), as shown in Table \ref{Table:frb_table}. In addition, we note that FRB 20201124A listed in Table \ref{Table:frb_table} is a repeating FRB, and suggest that quasi-periodic substructures in repeating FRBs may provide valuable insights for periodicity searches in these repeaters. However, as shown in Figure \ref{fig:pp_relation}, the predicted rotational periods ($< 1\,\text{s}$) of the host stars of FRBs 20200120E and 20201020A show statistically are much smaller than typical magnetar periods, which may suggest a non-unified origin scenario for FRBs or this method of inferring the host star's rotational period through measurements of quasi-periodic microstructure is unreliable. We look forward to future studies confirming that the host stars of these FRBs are consistent with our results, and anticipate detecting quasi-periodic substructures in more FRBs, particularly in repeating FRBs.

\begin{table}[]
\caption{Dataset for fitting the relation between the rotation periods and periods of quasi-periodic substructures}
\label{Table:dataset_table}
\centering
\begin{tabular}{ccccc}
\hline
Name                            & Type      & $P$          & $P_{\mu}$       & Reference \\
                                &           & ($\text{s}$) & ($\text{ms}$)   &          \\
\hline 
PSR J1622$-$4950                & Magnetar  & 4.326          & 10.0(0.2)       &\citep{2024NatAs...8..230K}\\
XTE J1810$-$197                 & Magnetar  & 5.540          & 12.0(0.06)      &\citep{2024NatAs...8..230K}\\
Swift J1818.0$-$1607            & Magnetar  & 1.364          & 2.24(0.04)      &\citep{2024NatAs...8..230K}\\
GLEAM-X J162759.5$-$523504.3    & Magnetar  & 1091.169       & 6110(290)       &\citep{2024NatAs...8..230K}\\
J0437$-$4715                    & MSP       & 0.005757       & 0.004(0.00079)
&\citep{2016ApJ...833L..10D}\\
J2145$-$0750                    & MSP       & 0.016052       &0.01536(0.00177)
&\citep{2016ApJ...833L..10D}\\
J1022+1001                      & MSP       & 0.016453       & $0.0149^{+0.0052}_{-0.0028}$
&\citep{2022MNRAS.513.4037L}\\
J2145$-$0750                    & MSP       & 0.016052       & $0.0178^{+0.0027}_{-0.0032}$
&\citep{2022MNRAS.513.4037L}\\
J1744$-$1134                    & MSP       & 0.004075       & $0.006^{+0.0024}_{-0.0008}$
&\citep{2022MNRAS.513.4037L}\\
J2144$-$3933                    & Long-rotation-period Pulsar& 8.510  & 11.8(5.83)      &\citep{2020MNRAS.492.2468M}\\
J0901$-$4046                    & Long-rotation-period Pulsar& 75.886 & $100(29)^{*}$   &\citep{2022NatAs...6..828C}\\
J0139+3336                      & RRAT      & 1.248          & 0.91                     &\citep{2024MNRAS.528.1213D}\\
J1913+1330                      & RRAT      & 0.923          & 1.56                     &\citep{2024MNRAS.527.4129Z}\\
J1913+1330                      & RRAT      & 0.923          & $0.26^{+0.04}_{-0.01}$   &\citep{2025MNRAS.tmp..545T}\\
B0301+19($-2.2^{\circ}\textendash2.5^{\circ}$)  &Normal Pulsar& 1.387 & 1.19(0.44)
&\citep{2015ApJ...806..236M}\\
B0525+21($-2.85^{\circ}\textendash2.86^{\circ}$)&Normal Pulsar& 3.745 & 3.57(1.76)
&\citep{2015ApJ...806..236M}\\
J0546+2441                      & Normal Pulsar& 2.843       & 2.38(0.71)
&\citep{2015ApJ...806..236M}\\
J0659+1414                      & Normal Pulsar& 0.384       & 0.42(0.09)
&\citep{2015ApJ...806..236M}\\
J0754+3231                      & Normal Pulsar& 1.442       & 2.10(0.98)
&\citep{2015ApJ...806..236M}\\
J0826+2637                      & Normal Pulsar& 0.530       & 0.72(0.45)
&\citep{2015ApJ...806..236M}\\
J0837+0610($-1.32^{\circ}\textendash1.46^{\circ}$)&Normal Pulsar&1.275& 0.72(0.18)
&\citep{2015ApJ...806..236M}\\
J0953+0755                      & Normal Pulsar& 0.253       & 0.63(0.22)
&\citep{2015ApJ...806..236M}\\
J1136+1551($-1.47^{\circ}\textendash3.39^{\circ}$)&Normal Pulsar&1.187& 0.83(0.44)
&\citep{2015ApJ...806..236M}\\
J1239+2453                      & Normal Pulsar& 1.382       & 2.16(1.16)
&\citep{2015ApJ...806..236M}\\
J1740+1000                      & Normal Pulsar& 0.154       & 0.48(0.18)
&\citep{2015ApJ...806..236M}\\
J1740+1311                      & Normal Pulsar& 0.803       & 1.80(0.80)
&\citep{2015ApJ...806..236M}\\
J1910+0714                      & Normal Pulsar& 2.712       & 5.36(2.21)
&\citep{2015ApJ...806..236M}\\
J1912+2104                      & Normal Pulsar& 2.233       & 4.40(2.12)
&\citep{2015ApJ...806..236M}\\
J1921+2153($-2.45^{\circ}\textendash3.10^{\circ}$)&Normal Pulsar&1.337& 1.31(0.62)
&\citep{2015ApJ...806..236M}\\
J1932+1059                      & Normal Pulsar& 0.226       & 0.48(0.18)
&\citep{2015ApJ...806..236M}\\
J1946+1805                      & Normal Pulsar& 0.440       & 0.71(0.18)
&\citep{2015ApJ...806..236M}\\
J2004+3137                      & Normal Pulsar& 2.111       & 1.25(0.53)
&\citep{2015ApJ...806..236M}\\
J2018+2839                      & Normal Pulsar& 0.557       & 0.66(0.18)
&\citep{2015ApJ...806..236M}\\
J2022+2854($-3.78^{\circ}\textendash1.21^{\circ}$)&Normal Pulsar&0.343& 0.48(0.18)
&\citep{2015ApJ...806..236M}\\
J2037+1942                      & Normal Pulsar& 2.074       & 3.10(1.50)
&\citep{2015ApJ...806..236M}\\
J2113+2754                      & Normal Pulsar& 1.202       & 1.13(0.09)
&\citep{2015ApJ...806..236M}\\
J2317+2149                      & Normal Pulsar& 1.444       & 1.01(0.09)
&\citep{2015ApJ...806..236M}\\
J0627+0706(MP)                  & Normal Pulsar& 0.476       & $0.82^{+0.17}_{-0.12}$
&This Work\\
J0826+2637(MP)                  & Normal Pulsar& 0.531       & $0.49^{+0.13}_{-0.10}$
&This Work\\
J0953+0755(MP)                  & Normal Pulsar& 0.253       & $0.50^{+0.15}_{-0.10}$
&This Work\\
J1946+1905(MP)                  & Normal Pulsar& 0.441       & $0.75^{+0.16}_{-0.16}$
&This Work\\
\hline
\end{tabular}
\tablecomments{Columns 1 and 2 list the name and the type of stars, respectively. Column 3 and 4 provide periods of the rotation periods and the characteristic periods of quasi-periodic substructure, respectively. Reference are list in the last column. The characteristic period marked with asterisk (*) for PSR J0901$-$4046 indicates rough estimate based on the method described in the main text. The data cited from M15 was obtained using smoothing bandwidths of $0.075\times N_{bin}$, and only that from the stronger integrated profile component was considered for the pulsars (marked longitude ranges in the table) whose microstructure in different components were investigated (see M15 for details).}
\end{table}

\begin{table}[]
\caption{Information about Quasi-periodic substructures in FRBs}
\label{Table:frb_table}
\centering
\begin{tabular}{cccc}
\hline
NAME      & $P_{\mu}$  & Reference & $P_{F}$ \\
          &($m\text{s}$)&           & ($\text{s}$)   \\
\hline 
20191221A & $216.8\pm0.1$ & \cite{2022Natur.607..256C}& $122\pm22$\\
20200120E & 0.0023   & \cite{2021ApJ...919L...6M}& $0.0023\pm0.0005$\\
20201020A & 0.411 & \cite{2023AA...678A.149P}& $0.328\pm0.029$\\
20201124A & $6.409^{+3.327*}_{-1.540}$ & \cite{2022RAA....22l4004N}& $4.37^{+3.2}_{-1.4}$\\
20210206A & $2.8\pm0.1$   & \cite{2022Natur.607..256C}& $2.02\pm0.18$\\
20210213A & $10.7\pm0.1$  & \cite{2022Natur.607..256C}& $7.07\pm0.76$\\
\hline
\end{tabular}
\tablecomments{Column 1 lists the name of FRBs. Column 2 provides periods of quasi-periodic substructure in these FRBs. Reference are list in Column 3. The predicted rotational periods of the host stars for these FRBs are listed in the last column. It should be noted that the characteristic period of quasi-periodic substructures in FRB 20201124A (marked with a superscript *) are represented by their median values, with uncertainties determined from the upper and lower quartiles.}
\end{table}

\subsection{Morphological Analysis of Single-pulses}
During the data processing, we noticed some interesting single-pulse morphologies. In this section, we focus on analyzing the morphology of single-pulses, which exhibit significant variability both across different epochs and within individual epochs, while certain characteristic features remain persistent. Following the classification scheme proposed by \cite{2022NatAs...6..828C}, single pulses can be categorized into seven distinct types: normal, quasi-periodic, spiky, double-peaked, partially nulling, split-peak, and triple-peaked, as illustrated in Figure 2 of \cite{2022NatAs...6..828C}. Notably, partially nulling pulses represent a distinctive phenomenon marked by an abrupt cessation of pulse intensity at specific phases. However, we emphasize that this classification is not strictly rigid or mutually exclusive. In particular, we observed that quasi-periodic features can coexist with partially nulling pulses as well as with pulses exhibiting multiple peaks (e.g., double-peaked or triple-peaked pulses), suggesting a continuum of pulse morphologies influenced by complex underlying physical processes.

To further explore the diversity of pulse characteristics, we also performed a classification based on the presence or absence of low-frequency envelope features. In the following subsections, we analyzed the morphological properties of single pulses for each pulsar in our sample. This investigation aims to uncover patterns and variations in pulse morphology, which may provide critical insights into the physical mechanisms governing pulse emission processes.

\begin{figure}[ht!]
\centering
\includegraphics[scale=0.2475]{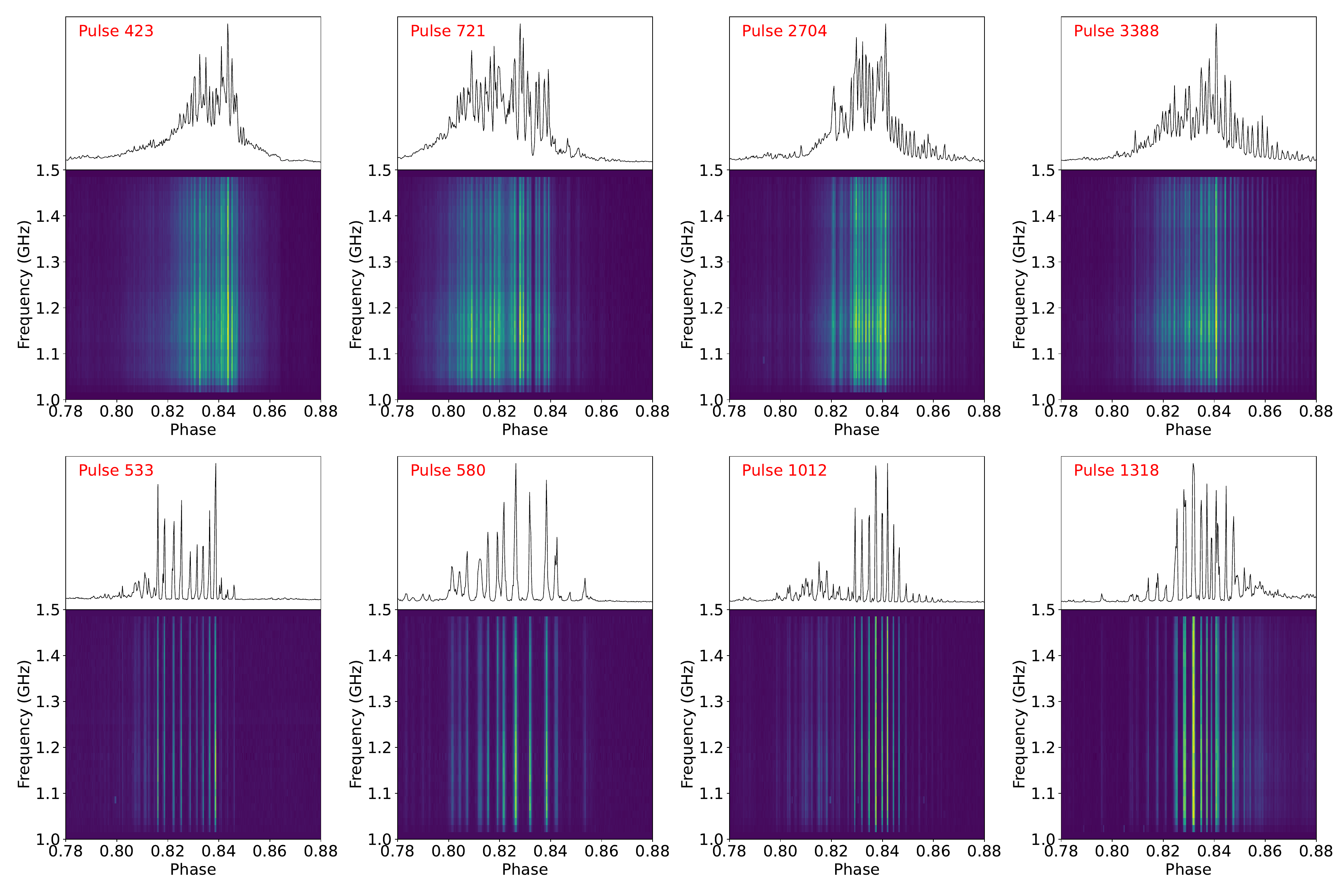}
\caption{Gallery of quasi-periodic microstructure in the MP of PSR J0953+0755. Each panel features a pulse profile (top) and a phase-frequency intensity map (bottom), with data downsampled to 32 frequency channels.
\label{fig:0950+08_mp_period}
}
\end{figure}
\begin{figure}[hb!]
\centering
\includegraphics[scale=0.2475]{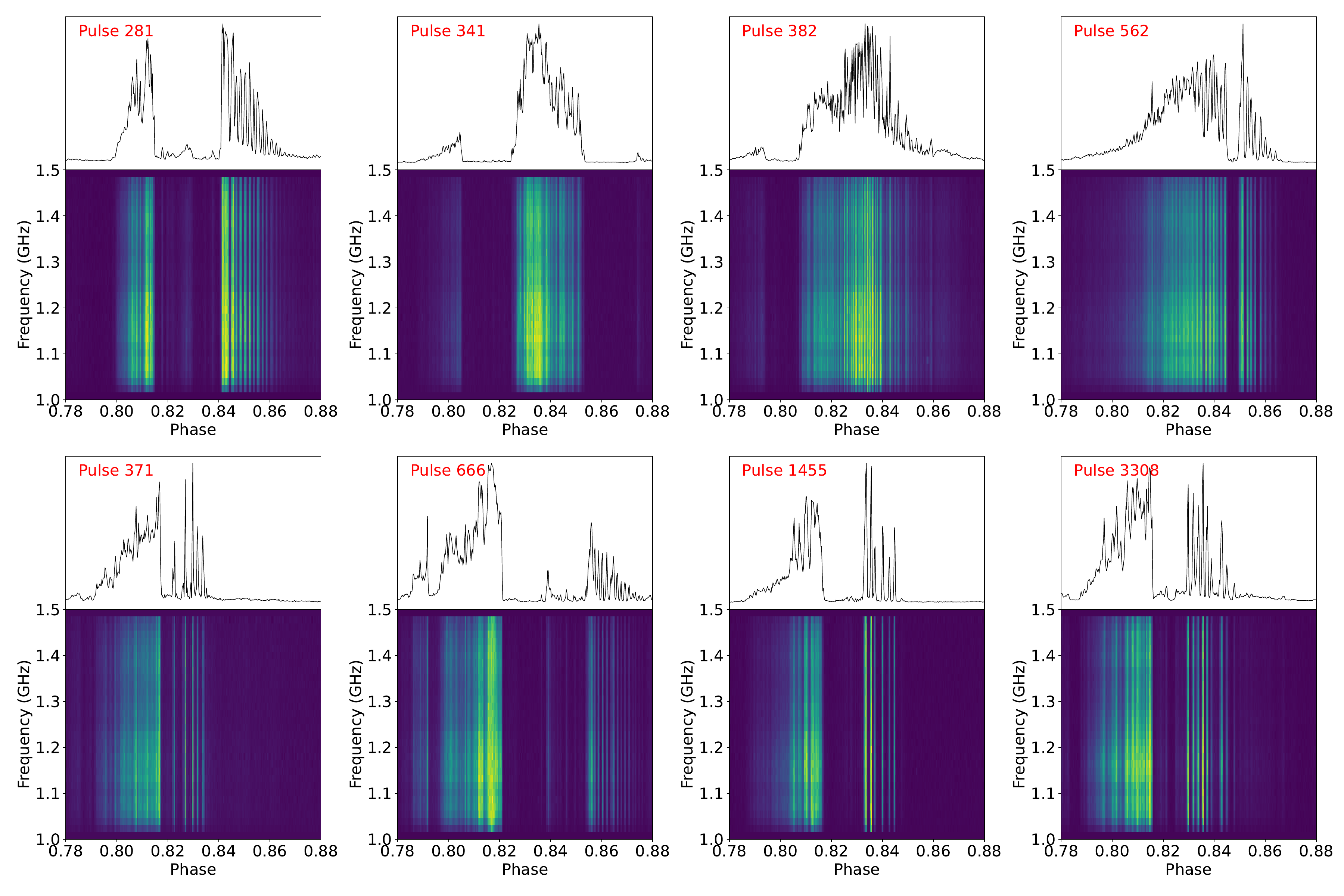}
\caption{Gallery of partially nulling pulses in the MP of PSR J0953+0755. Each panel features a pulse profile (top) and a phase-frequency intensity map (bottom), with data downsampled to 32 frequency channels.
\label{fig:0950+08_mp_nulling}
}
\end{figure}
\begin{figure}[ht!]
\centering
\includegraphics[scale=0.2475]{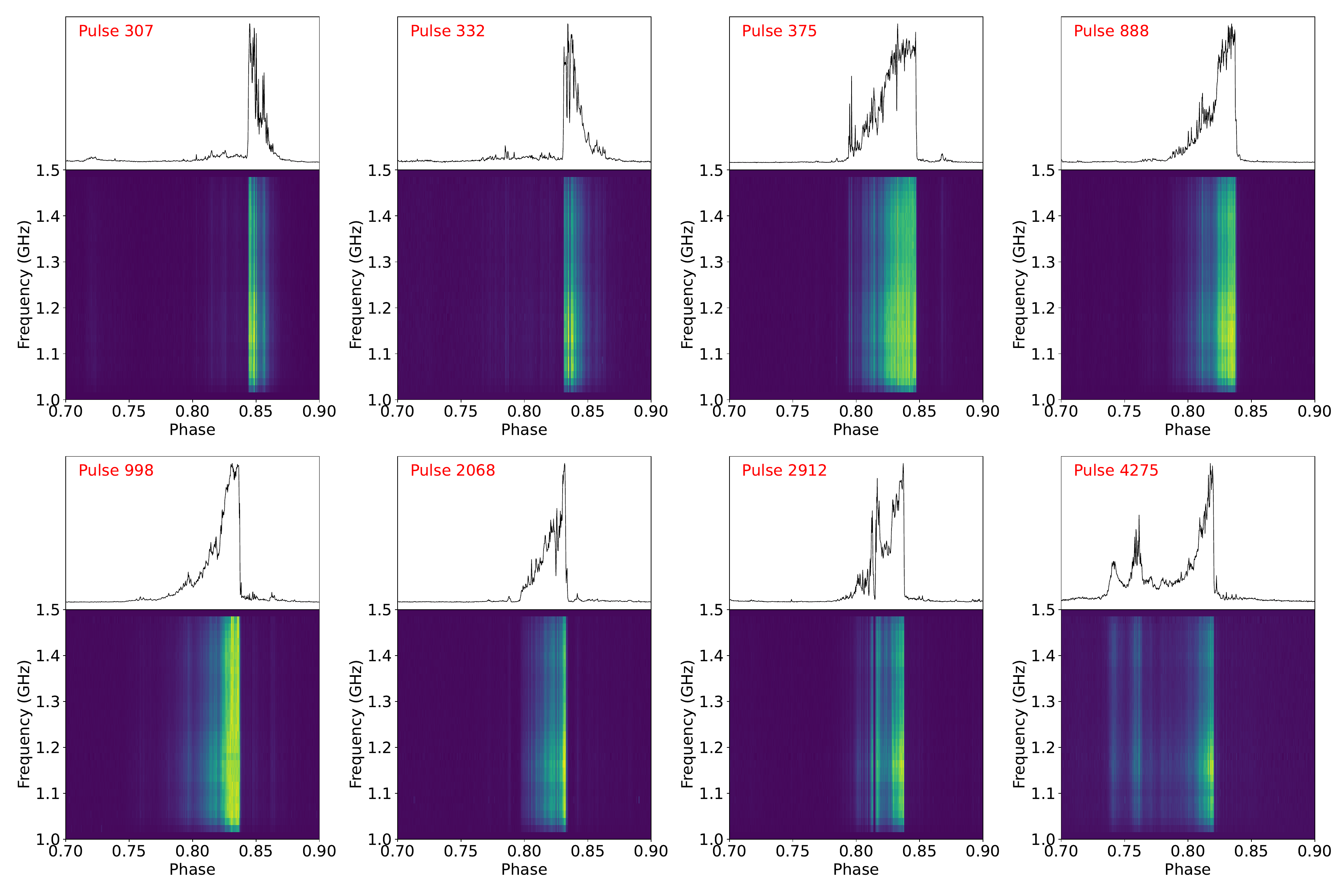}
\caption{Gallery of pulses exhibiting rapid intensity increase or sharp decay in the MP of PSR J0953+0755. Each panel features a pulse profile (top) and a phase-frequency intensity map (bottom), with data downsampled to 32 frequency channels.
\label{fig:0950+08_mp_rapid}
}
\end{figure}
\begin{figure}[hb!]
\centering
\includegraphics[scale=0.2475]{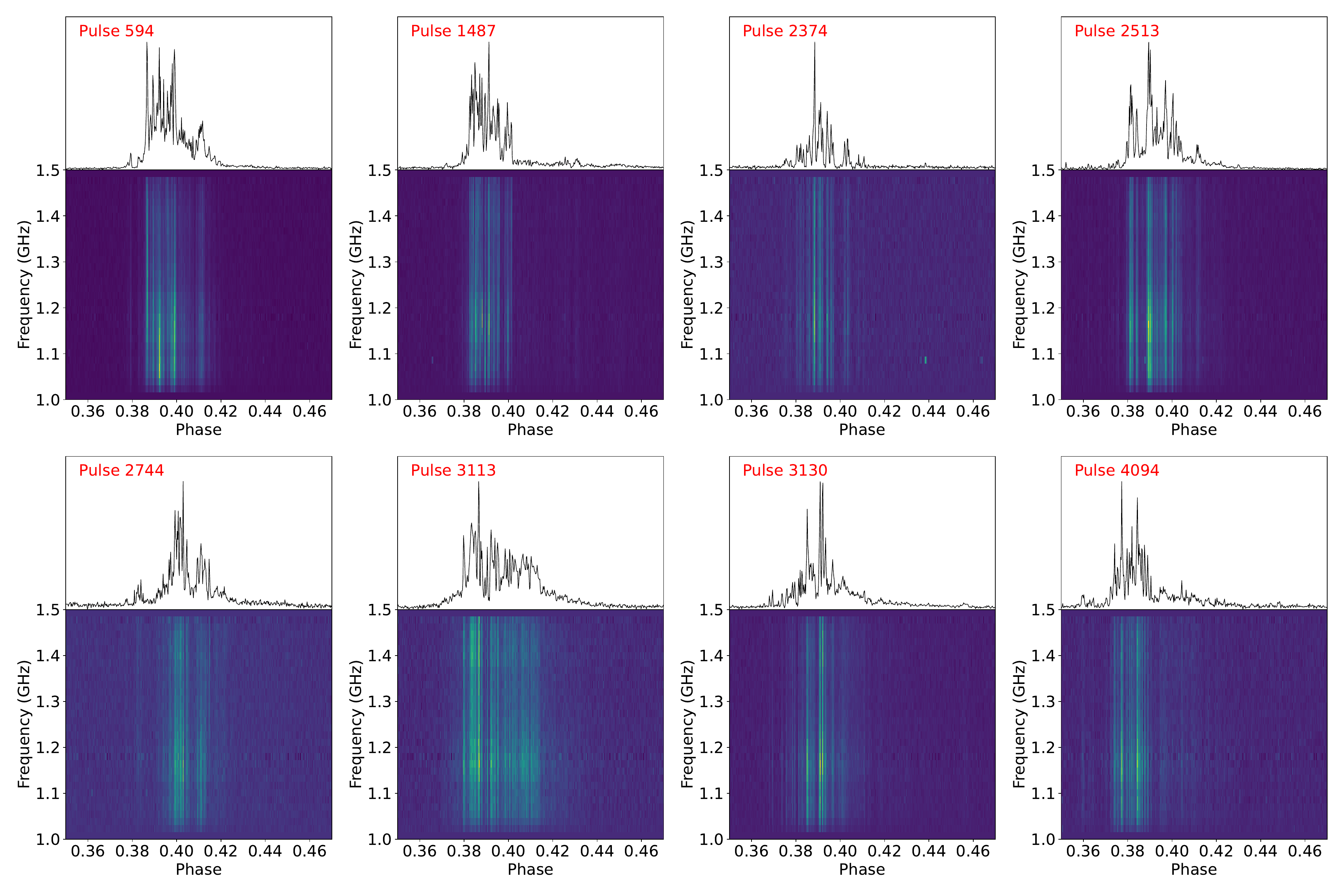}
\caption{Gallery of quasi-periodic microstructure in the IP of PSR J0953+0755. Each panel features a pulse profile (top) and a phase-frequency intensity map (bottom), with data downsampled to 32 frequency channels.
\label{fig:0950+08_ip_period}
}
\end{figure}
\begin{figure}[ht!]
\centering
\includegraphics[scale=0.2475]{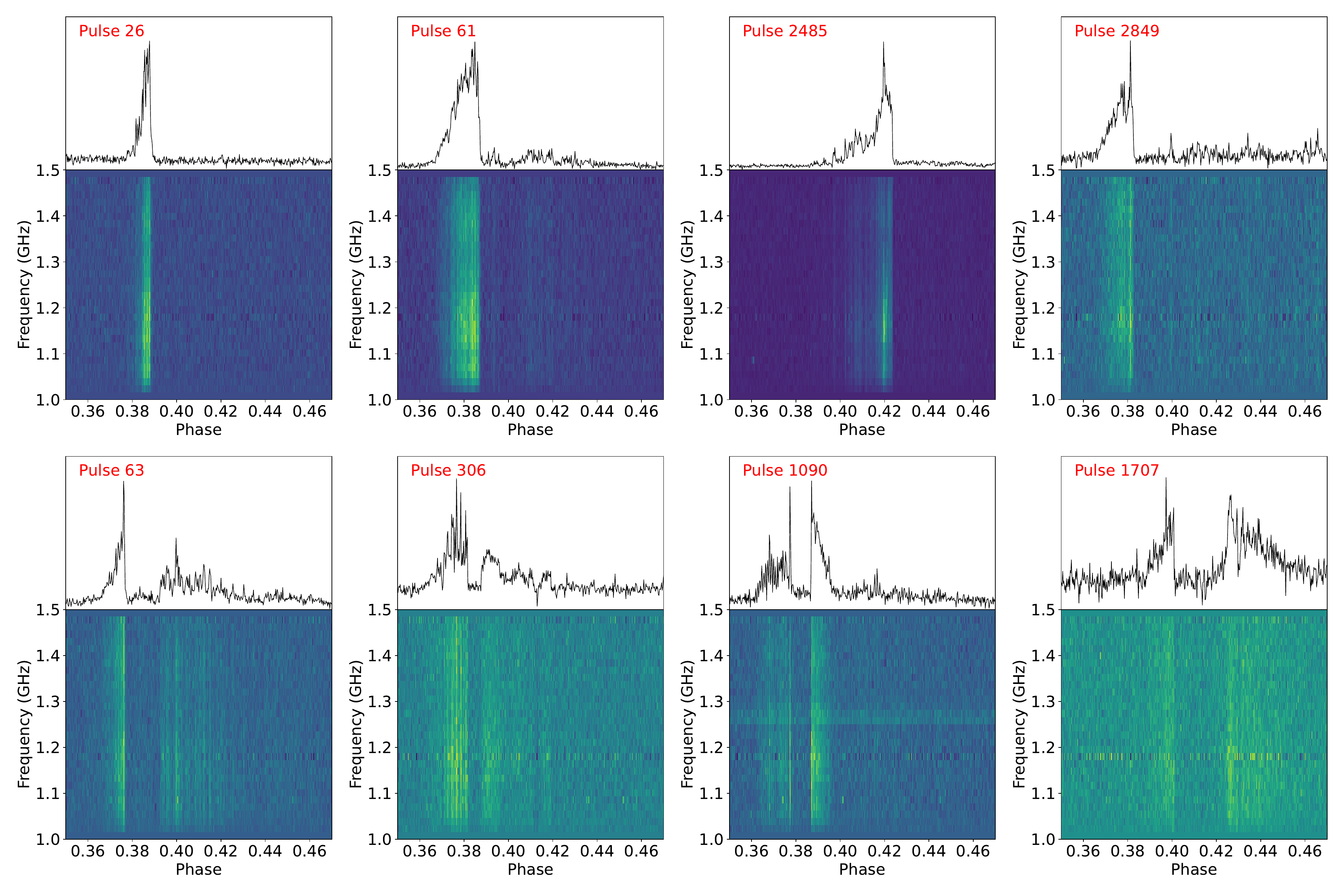}
\caption{Gallery of partially nulling pulses and pulses exhibiting rapid intensity decay in the IP of PSR J0953+0755. Each panel features a pulse profile (top) and a phase-frequency intensity map (bottom), with data downsampled to 32 frequency channels.
\label{fig:0950+08_ip_null_rapid}
}
\end{figure}

\textbf{PSR J0953+0755:} In the MP of this pulsar, we observed various single-pulse morphologies, including those with low-frequency envelopes (top panels of Figure \ref{fig:0950+08_mp_period}) and those without (bottom panels of Figure \ref{fig:0950+08_mp_period}). Notably, our analysis revealed not only partially nulling pulses but also instances where partial-nulling structures and quasi-periodic microstructure coexist within individual pulses, as illustrated in the panels of Figure \ref{fig:0950+08_mp_nulling}. Additionally, we find that certain nulling pulses maintain consistent pulse profiles before and after the nulling phase, as depicted in the top four panels of Figure \ref{fig:0950+08_mp_nulling}. In contrast, other partially nulling pulses undergo significant changes in their pulse profiles following the nulling phase, as demonstrated in the bottom four panels of Figure \ref{fig:0950+08_mp_nulling}, in which, these post-nulling pulse components display quasi-periodic microstructure without low-frequency envelopes.

Beyond the partially nulling pulses characterized by an abrupt intensity drop, we identified a subset of pulses exhibiting rapid intensity increases (Pulses 307 and 332) or sharp decays, as illustrated in Figure \ref{fig:0950+08_mp_rapid}. Notably, the abrupt increases or decreases in the intensity of these single pulses shown in this figure all occur near a phase of 0.83, which may imply an unrecognized physical mechanism. However, whether this phenomenon is a coincidence still needs to be verified through further research.

In the IP of this pulsar, almost all pulses exhibiting quasi-periodic microstructure are superimposed on low-frequency envelopes, as demonstrated in Figure \ref{fig:0950+08_ip_period}. Furthermore, our analysis of the IP has also revealed the presence of both partially nulling pulses and pulses exhibiting rapid intensity increases, as shown in Figure \ref{fig:0950+08_ip_null_rapid}.

\textbf{PSRs J0627+0706, J0826+2637 and J1946+1805:} In these three pulsars, we did not detect any quasi-periodic microstructure without low-frequency envelopes. Furthermore, no significant partially nulling pulses or pulses exhibiting rapid intensity increase or sharp decay were observed. This suggests that partially nulling pulses, quasi-periodic microstructure without low-frequency envelopes, and pulses with rapid intensity variations may not be common phenomena within the pulsar population.

We propose that partial nulling phenomena may arise from the line of sight sweeping through gaps between sub-beamings. The phase inconsistency in partial nulling may be explained by variations in sub-beamings caused by fluctuations in plasma density. Note that although we report these interesting single-pulse morphologies in this paper, they are not the focus here. Therefore, we refrain from conducting a systematic analysis in this study and will pursue in-depth investigations in subsequent research.

\section{Discussion} \label{sec:discussion}
\subsection{Physical Origin of Microstructure}
The physical origin of pulsar microstructure is an important question in understanding the mechanisms of radio emission of pulsars. Current theories can be broadly categorized into geometric beaming effects, temporal/radial modulation effects, plasma instabilities, and coherent radiation mechanisms, as detailed below:

\begin{table}[]
\caption{The Lorentz Factors ($\gamma$) of Pulsars Derived from Microstructure}
\label{Table:gamma_table}
\centering
\begin{tabular}{ccccccc}
\hline
NAME         & P           &$\tau_{\mu}$           & Reference                 &$\zeta$                  &Reference                  & $\gamma$\\
             & (s)         &($m\text{s}$)          &                           &                         &                           &         \\
\hline 
J1022+1001   & 0.01645     &$0.009\pm0.001$        &\cite{2022MNRAS.513.4037L} &7.9(1.4)                 &\cite{2024arXiv241205452F} & 2116.5  \\
J1932+1059   & 0.226       &$0.24(0.09)^{*}$       &\cite{2015ApJ...806..236M} &$54.32^{+2.21}_{-2.28}$  &\cite{2025arXiv250313824S} & 184.5   \\
J2144$-$3933 & 8.510       &$5.9(2.91)$            &\cite{2020MNRAS.492.2468M} &72.03                    &\cite{2020MNRAS.492.2468M} & 241.3   \\
J0953+0755   & 0.253       &$0.25^{+0.06}_{-0.00}$ &This Work                  &$178.49^{+4.68}_{-3.90}$ &\cite{2025arXiv250313824S} & 6112.2  \\
J1946+1805   & 0.441       &$0.43^{+0.00}_{-0.11}$ &This Work                  &$170.25^{+4.83}_{-5.74}$ &\cite{2025arXiv250313824S} & 963.8   \\

\hline
\end{tabular}
\tablecomments{Columns 1 and 2 list the pulsar name and its rotation period, respectively. Columns 3 and 4 present the characteristic timescales of microstructure and their corresponding references. Columns 5 and 6 provide the $\zeta$ values and their associated references, respectively. The derived Lorentz factors ($\gamma$) are listed in the final column. $\tau_{\mu}$ of PSR J1932+1059 is an approximate estimate obtained by halving the characteristic period of microstructure reported in M15.}
\end{table}

\textbf{Angular Beaming and Relativistic Particle Flux Tubes:} A leading paradigm posits that microstructure arises from narrow flux tubes of relativistic charged particles streaming along curved magnetic field lines, whose radiation is beamed in their direction of motion. As the pulsar rotates, these beams sweep across the observer’s line of sight, with the width of the microstructure reflecting the width of the angular beam \citep{1977MNRAS.179..311B, 1980A&A....87..282B}. A point source moving with relativistic energy emits radiation that is beamed in the direction of its motion. If assuming that the fine structure is a consequence of a flux tube, which consists of such point sources, sweeping across our line of sight, then, under the assumption that the flux tube is infinitely narrow, we can deduce a lower bound for the particle energy from the width of the micropulses \citep{1998A&A...332..111L}:
\begin{equation}
\gamma = \frac{P}{2\pi \tau_{\mu} \cdot \sin(\zeta)}
\end{equation}
where $\gamma$ is the Lorentz factor of the particles, $P$ is the rotation period of pulsar, $\tau_{\mu}$ is the characteristic timescale of microstructure, and $\zeta=\alpha+\beta$ is the angle between the line of sight and the rotation axis. Some researchers have argued that $\gamma$ should not significantly exceed a value of 100 \citep{1993MNRAS.264..940A}. However, derived $\gamma$ values (from 240 to 520 \citep{1998A&A...332..111L}) often exceed theoretical expectations ($\gamma\leq100$) from simpler curvature radiation models \citep{1982ApJ...252..337D}. We also calculated the Lorentz factor for some pulsars. As shown in Table \ref{Table:gamma_table}, all these $\gamma$ are greater than 100. This significant discrepancy may suggest that such a model may not adequately account for the observed microstructure or the pulsar's radiation may not conform to simple coherent curvature radiation models. Assuming that the $\gamma$ is low, the next interpretation for the microstructure width is more likely to hold.

\textbf{Temporal Modulation:} Alternative models attribute microstructure to temporal or radial modulations within the magnetosphere. Temporal modulation via nonlinear plasma instabilities, such as two-stream instability, generates coherent radiation from charged solitons that excite extraordinary (X) and ordinary (O) modes in electron-positron plasmas \citep{2000ApJ...544.1081M}. The absence of circular polarization sign reversals in observations supports this scenario, as plasma propagation effects disrupt phase coherence between modes, unlike vacuum curvature radiation from single particles \citep{2015ApJ...806..236M}. Radial modulation models link microstructure duration to the thickness of emitting structures (e.g. plasma clouds), with radial lengths on the order of $c\cdot P_{\mu}$ (hundreds of kilometers for nanosecond pulses, as in the Crab pulsar; \cite{2003Natur.422..141H}). Interestingly, \cite{2022ApJ...933..231T, 2022ApJ...933..232T} link the radio emission of rotating neutron stars with slow tearing instabilities caused by uneven twisting in the open pulsar circuit. In this theory, radio emission comes from coherent curvature radiation generated by Cherenkov-like instabilities in current-carrying Alfvén waves within thin sheets of relativistic particle flow \citep{2022ApJ...933..232T}. The model also predicts that directed radio radiation arises from clusters of charged particles, with quasi-periodic microstructure emerging naturally from these physical processes.
The observed timescale of the microstructure, influenced by relativistic beaming, leads to the derived relation $\tau_{\mu}\approx1/(\Omega\gamma_{eff})$, which matches the observed relationship $P_{\mu}\approx10^{-3}P$ \citep{2022ApJ...933..232T}. This result suggests that all radio-emitting neutron stars might share the same effective Lorentz factor: $\gamma_{eff}\approx200$. It is worth noting that the physical conditions associated with slow tearing instability are more significantly influenced by the local magnetic field strength. Additionally, the magnetic field of pulsars may not be a simple dipole field, which seemingly explains why there is no obvious correlation between the microstructure and the pulsar surface magnetic field derived under the dipole field assumption in M15.

\textbf{Observational Constraints and Open Questions:} Key evidence favoring magnetospheric origins includes the consistency of Stokes parameter periodicities (I, V, L), ruling out single-particle radiation \citep{2015ApJ...806..236M}, and frequency-dependent width scaling consistent with radius-frequency mapping \citep{1976ApJ...208..944C}. However, unresolved issues persist: the discrepancy between beaming-model $\gamma$ factors and theoretical limits and the role of polar cap asymmetry in generating component-specific microstructure (e.g. in PSR B1133+16; \cite{2002A&A...396..171P}). It is worth noting that previous studies reported differences in the microstructure of different components \citep{2002A&A...396..171P}, but the results presented in M15 indicate that there are no significant differences in the microstructure among different components. In this work, our results indicate that the $\tau_{\mu}$ and $P_{\mu}$ of microstructure across components appear consistent within measurement errors for PSR J0627+0706, but microstructure in IP are relatively smaller than those in MP for PSR J0953+0755. Whether the microstructure of different components are consistent and what factors influence such differences require more observational evidence for further research. More importantly, how can we determine whether microstructures result from a combination of temporal modulation and angular beam models? Or, what level of 
time resolution is required in observations to conclude that the observed microstructures are dominated by temporal modulation? These questions still require further investigation.

In summary, while angular beaming and plasma modulation models provide complementary frameworks, integrating high-resolution polarization and multi-frequency data, especially for millisecond pulsars and magnetars, will be pivotal to resolving the interplay between particle acceleration, radiation geometry, and propagation effects in shaping pulsar microstructure.

\subsection{Microstructure as a Probe for Correlations Between MP and IP}
Based on the theoretical hypothesis that microstructures are formed by temporal modulation within magnetic flux tubes, and considering the relative stability of the time scales and characteristic periods of microstructures within the same magnetic flux tube, we can determine whether the interpulse (IP) and main pulse (MP) of pulsars originate from the same magnetic flux tube by comparing the similarities and differences of their microstructure. Generally, if the pulse signals originate from the same magnetic flux tube, their polarization properties maybe similar. In contrast, signals from different magnetic regions exhibit abrupt jumps in polarization phase and do not follow the S-shaped wings. \cite{2025arXiv250313824S} conducted an in-depth study on the MP and IP of pulsars. Their results show that the observational data of PSR J0953+0755 conform to the characteristics of the Rotating Vector Model (RVM), suggesting that its MP and IP may originate from the same magnetic pole. Meanwhile, this pulsar shows a significant jump in the polarization position angle. A further detailed comparative analysis of the microstructures of the MP and IP of this pulsar reveals that the microstructures in the IP are more refined in size compared to those in the MP. Combining with the above-mentioned theoretical framework, this correlation in microstructure size provides strong observational evidence that the MP and IP of J0953+0755 originate from different magnetic flux tubes of the same magnetic pole.
For PSR J0627+0706, its polarization properties deviate from the predictions of the RVM model, the polarization position angle of both MP and IP component exhibit abrupt jumps \citep{2025arXiv250313824S}, the separation between the MP and IP components is approximately half a rotation period, and the microstructure in this pulsar show no significant differences between MP and IP. Thus, we conclude that the MP and IP components originate from two distinct magnetic poles, and the MP might share the same magnetic flux tube with part of the IP component. The pulse profile of PSR J0826+2637 is intriguing, composed of a MP, a subsequent post-cursor (PC), and an IP. Results from \cite{2025arXiv250313824S} show that the polarization position angle of this pulsar exhibits a distinct jump in the phase between the MP and PC, and the phase separation between the MP and IP is approximately half a rotation period. Thus, we propose that the MP and IP originate from different magnetic poles and might share the same magnetic flux tube (we emphasize that this is not a robust conclusion due to the lack of comparative analysis in microstructure). Some studies suggest that pulsars may not strictly follow simple dipolar fields \citep{2019MNRAS.489.4589A}. Therefore, a plausible explanation is that the PC component originates from local strong magnetic field regions other than the dipolar field.

\section{Conclusion} \label{sec:conclusion}
In this paper, we studied microstructure in four interpulse pulsars (PSRs J0953+0755 (B0950+08), J0627+0706, J0826+2637 (B0823+26) and J1946+1805 (B1944+17)) with FAST. The main conclusions are as follows:
\begin{enumerate}
\item For the first time, we detected the quasi-periodic microstructure in the interpulse of pulsars.
\item We found that microstructure, even quasi-periodic microstructure, seems to exit in the mainpulse and interpulse of all pulsars, except for interpulse of PSRs J0826+2637 and J1946+1805, and obtained the statistical results of characteristic timescales and characteristic periods of these microstructure.
\item The microstructure in mainpulse and interpulse were compared, and our results show that ,for PSR J0627+0706, the characteristic timescales and characteristic periods of quasi-periodic microstructure between MP and IP are generally consistent within error ranges. However, for PSR J0953+0755, microstructure in IP may be smaller than those in MP.
\item  The relationship between the characteristic period of microstructure and the rotation period in neutron star populations was reconfirmed: $P_{\mu}(\text{ms})=(1.337\pm0.114)\times P(\text{s})^{(1.063\pm0.038)}$.
\end{enumerate}

\section{Acknowledgments}
This work was supported by the Major Science and Technology Program of Xinjiang Uygur Autonomous Region (No.2022A03013-4), the Guizhou Provincial Basic Research Program (Natural Science) (No. Qiankehejichu-MS(2025)263), the Guizhou Provincial Science and Technology Projects (Nos. QKHFQ[2023]003, QKHPTRC-ZDSYS[2023]003, QKHFQ[2024]001-1), Guizhou Provincial Scientific and Technological Program$-$the Postdoctoral Research Station of the Guizhou Radio Astronomical Observatory (No.QKHPTRC[2021] Postdoctoral Research Station-001), the National Natural Science Foundation of China (No. 12041304), the Project funded by China Postdoctoral Science Foundation(No. 2023M743517) and the Natural Science Basic Research Program of Shaanxi (Program No. 2024JC-YBQN-0036). This work made use of the data from FAST (Five-hundred-meter Aperture Spherical radio Telescope).  FAST is a Chinese national mega-science facility, operated by National Astronomical Observatories, Chinese Academy of Sciences.\

\vspace{5mm}

\software{PSRCHIVE \citep{2004PASA...21..302H},  
          DSPSR \citep{2011PASA...28....1V}, 
          PSRCAT \citep{2005AJ....129.1993M}
  \texttt{NumPy}\footnote{https://numpy.org/} \citep{harris2020array}, \texttt{Matplotlib}\footnote{https://matplotlib.org/} \citep{Hunter:2007} and \texttt{SciPy}\footnote{https://scipy.org/} \citep{2020SciPy-NMeth}.
  }

\bibliography{sample631}{}
\bibliographystyle{aasjournal}



\end{document}